\documentclass[twocolumn,amsmath]{revtex4}
\usepackage{graphicx} 
\usepackage{braket}  
\usepackage{amssymb}  
\usepackage{bm}
\usepackage{xcolor}
\usepackage{enumitem}
\usepackage{mathrsfs}

\newcommand{\ketbra}[2]{\mathinner{|{#1}\rangle \! \langle{#2}|}}

\newcommand{\sref}[1]{Sec.~\ref{#1}}
\newcommand{\aref}[1]{Appendix~\ref{#1}}
\newcommand{\eref}[1]{Eq.~(\ref{#1})}
\newcommand{\erefs}[1]{Eqs.~(\ref{#1})}
\newcommand{\fref}[1]{Fig.~\ref{#1}}
\newcommand{\frefs}[1]{Figs.~\ref{#1}}
\newcommand{\rcite}[1]{Ref.~\cite{#1}}
\newcommand{\rcites}[1]{Refs.~\cite{#1}}
\newcommand{\tref}[1]{Table~\ref{#1}}

\usepackage{hyperref}
\hypersetup{
colorlinks=true,
linkcolor=blue,
filecolor=blue,
citecolor=blue
}

\begin{document}

\title{Testing bath correlation functions for open quantum dynamics simulations}

\author{Masaaki Tokieda}
\email{tokieda.masaaki.4e@kyoto-u.ac.jp}
\affiliation{Department of Chemistry, Graduate School of Science, Kyoto University, Kyoto, Japan}

\begin{abstract}

Accurate simulations of thermalization in open quantum systems require a reliable representation of the bath correlation function (BCF). Numerical approaches, such as the hierarchical equations of motion and the pseudomode method, inherently approximate the BCF using a finite set of functions, which can impact simulation accuracy. In this work, we propose a practical and rigorous testing framework to assess the validity of approximate BCFs in open quantum dynamics simulations. Our approach employs a harmonic oscillator system, where the computed dynamics can be benchmarked against known exact solutions.
To enable practical testing, we make two key methodological advancements. First, we develop numerical techniques to efficiently evaluate these exact solutions across a wide range of BCFs, ensuring broad applicability. Second, we introduce a moment-based state representation that significantly simplifies computations by exploiting the Gaussian nature of the system.
Applications to a two-spin system and a transmon-resonator system demonstrate that our testing procedure provides error estimates that capture the qualitative trends observed in thermalization simulations.
Using this methodology, we assess the performance of recently proposed BCF construction methods, highlighting both their strengths and a notable challenge posed by sub-Ohmic spectral densities at finite temperatures.

\end{abstract}

\maketitle

\section{Introduction}

Open quantum dynamics describe the evolution of a quantum system interacting with its environment. Simulations of such dynamics serve various purposes.
One key objective, central focus of this article, is to provide a dynamical description of thermalization.
In the context of open quantum dynamics, thermalization refers to the relaxation of the open system toward a unique steady state corresponding to the reduced Gibbs state when coupled to a thermal environment (or heat bath).
Accurate descriptions of thermalization are essential for studying nonequilibrium phenomena \cite{SHT18,Tanimura20,LMR23,Jankovic23} and for advancing quantum thermodynamics \cite{HSAL11,Llobet18,TH20,KT24_1,KT24_2}.

The bath oscillator model, in which a collection of harmonic oscillators (the bath) interacts with a system via a linear interaction operator, is widely used to study open quantum dynamics \cite{FV63,Ullersma66,CL83,FLO88,Weiss08}.
When the bath is initially in a Gaussian state, its influence on the system is entirely characterized by the two-time correlation function of the interaction operator, called the bath correlation function (BCF).
The imaginary part of the BCF, which is independent of the bath state, encodes the bath structure, and its Fourier transform defines the spectral density.
For a thermal bath, the BCF satisfies the fluctuation-dissipation relation, which is a key ingredient for thermalization as in the classical Brownian motion \cite{Chandrasekhar43}.
Thermalization is generally expected, as supported by studies employing various approximations (see \rcite{TMCA22} and references therein), by results for a specific system with general spectral densities \cite{GSI88} and for general systems with a specific spectral density \cite{Tanimura14}, with potential extensions to more general settings \cite{Koyanagi}.

Due to the lack of analytic solutions for most practical systems, numerical simulations are necessary.
Several methods have been established, including
Hierarchical Equations Of Motion (HEOMs)
\cite{Tanimura20,TK89,YYLS04,IT05,Tanimura06},
pseudomode method
\cite{Imamoglu94, Garraway97, Tamascelli17, Smirne22},
QuasiAdiabatic propagator Path Integral \cite{MM95_1,MM95_2}
and its efficient tensor network implementation called Time-Evolving Matrix Product Operator
\cite{SKKKL18,JP19,OQuPy24,LTS24},
MultiConfiguration Time-Dependent Hartree
\cite{MCTDH,WT03},
the chain mapping technique called Time Evolving Density operator with Orthogonal Polynomials Algorithm \cite{Chin10,Prior10,Tamascelli19,MPSDynamics},
and stochastic methods \cite{SM99,Diosi98,HOPS}.
In this article, we mainly focus on HEOM or the pseudomode method, whose connection was recently elucidated in \rcites{Link22,Xu23,MT24}.
These methods require approximating the BCF as a finite sum of functions, each corresponding to an auxiliary bosonic mode.
Intuitively, one can view the system as being coupled to a finite set of dissipative auxiliary modes, whose structure is engineered to reproduce the bath influences encoded in the BCF.
While reducing the number of auxiliary modes is desirable for computational efficiency, an inaccurate BCF leads to unrealiable results.
In particular, failure to satisfy the fluctuation-dissipation relation can lead to poor reproduction of the thermalization behavior.
Such issues become increasingly severe for more complex BCFs, which are gaining attention as more realistic models of physical environments \cite{FDV20,UT21,NA24,Lorenzoni25}.

While the error in an approximated BCF can be readily estimated, evaluating its impact on simulation results is nontrivial.
Previous works derived upper bounds on the error in expectation values based on the BCF or spectral density error \cite{Mascherpa17,Huang24}.
However, these bounds grow exponentially with time and significantly overestimate the actual error once the system reaches a steady state.
Thus, they are unsuitable for assessing the error for thermalization simulations.
A more practical approach is to examine convergence with increasing BCF accuracy.
However, for complex BCFs or systems, this procedure becomes time-consuming, motivating the need for a simpler test.

To this end, we propose replacing the system with a simpler and exactly solvable one, while maintaining the same BCF.
We then compare the simulation results with the exact solutions to assess the validity of an approximate BCF.
A well-known exactly solvable model is a dephasing model, where the system and interaction Hamiltonians commute \cite{Breuer02}.
However, this model does not thermalize due to the presence of conserved quantities.
As an alternative, we consider a harmonic oscillator system, which admits exact solutions and is known to thermalize for general spectral densities \cite{GSI88}.
In \rcite{Tanimura15}, non-Markovian tests were introduced to assess the accuracy of numerical methods using this system (see also \rcite{Koyanagi24}).
Here, we use them to test the BCF.
Thus far, the applications of non-Markovian tests have been limited to simple spectral densities, where exact solutions can readily be obtained.
We also note that the infinite-dimensional nature of the oscillator Hilbert space poses computational challenges and introduces another source of error due to the truncation of the system state.

To enable practical testing, this work offers two key contributions.
First, we revisit the exact solutions in \rcite{GSI88} and develop efficient numerical evaluation methods in order to extend their applicability to general spectral densities.
Second, we introduce a system state representation based on ladder operator moments \cite{Link22,MT24}.
Owing to the Gaussian nature of the total system, both system and auxiliary states can be exactly truncated at low orders in this representation.
Compared to other state representations (e.g., Fock basis or Wigner function), this approach simplifies the testing procedure by reducing computational costs and eliminating the need to calibrate truncation levels.
In this regard, we note that the use of the harmonic oscillator system is advantageous over the dephasing model, which requires truncating auxiliary states.

The remainder of this article is organized as follows.
In \sref{method}, we describe the testing methodologies.
In \sref{analysis}, we compare the performance of recently proposed algorithms for efficiently constructing approximate BCFs.
In \sref{demo}, we demonstrate the testing procedure using a two-spin system and a transmon-resonator system and show its effectiveness.
Finally, \sref{conclusion} provides a summary and concluding remarks.

\section{Methodologies}
\label{method}

\subsection{Bath oscillator model}
\label{method_model}

We consider open quantum dynamics described by a bath oscillator model.
The evolution of the total density operator, $\rho_{\rm tot}$, is governed by
\begin{equation*}
    \frac{d}{dt}\rho_{\rm tot}(t) = - \frac{i}{\hbar} [H_{\rm tot}, \rho_{\rm tot}(t)],
\end{equation*}
with
\begin{equation}
    H_{\rm tot} = H_S + \sum_m \hbar \omega_m b_m^\dagger b_m + V_S \sum_m u_m (b_m + b_m^\dagger).
    \label{eq:method_Htot}
\end{equation}
In this model, the bath consists of harmonic oscillators with annihilation (creation) operators $b_m$ ($b_m^\dagger$) for the $m$th mode obeying the bosonic commutation relations $[b_m, b_n^\dagger] = \delta_{m,n}$, with the Kronecker delta $\delta_{m,n}$.
Each mode has a frequency $\omega_m$ and couples to the system with strength $u_m$.
We define the bath Hamiltonian and interaction operator as $H_B \equiv \sum_m \hbar \omega_m b_m^\dagger b_m$ and $X_B \equiv \sum_m u_m (b_m + b_m^\dagger)$.
The Hermitian operators $H_S$ and $V_S$ describe the system's internal dynamics and its interaction with the bath, respectively.
The bath interaction modifies the bare potential in $H_S$, which can be compensated by introducing a counter term \cite{CL83}
\begin{equation}
    H_S = H_{S,{\rm eff}} + \lambda V_S^2,
    \label{eq:method_CT}
\end{equation}
with the effective system Hamiltonian $H_{S,{{\rm eff}}}$ and $\lambda = \sum_m u_m^2/(\hbar \omega_m)$.

We focus on the evolution of the system's reduced density operator, defined as $\rho_S(t) \equiv {\rm tr}_B [\rho_{\rm tot}(t)]$, where ${\rm tr}_B$ denotes the trace over the bath.
Assuming an initially uncorrelated state $\rho_{\rm tot}(0) = \rho_S(0) \rho_B$ with a thermal bath $\rho_B = \exp( -\beta H_B) / {\rm tr}_B[\exp(-\beta H_B)]$, where $\beta$ is the inverse temperature, the evolution of $\rho_S(t)$ is given by \cite{FV63}
\begin{equation}
    \rho_S(t) = e^{- i H_S^\times t / \hbar} \, \mathcal{T} [\mathcal{M}(t)] \, \rho_S(0),
    \label{eq:method_rhoS}
\end{equation}
with
\begin{equation}
\begin{gathered}
    \mathcal{M}(t) =
    \exp \left[ - \frac{1}{\hbar} \int_0^t d\tau \int_0^\tau du \right. \\ V_S^I(\tau)^\times
    \left. \left\{ L(\tau-u) V_S^I(u)^L - L^*(\tau-u) V_S^I(u)^R \right\} \right] ,
\end{gathered}
\label{eq:method_M}
\end{equation}
where we introduce the chronological time-ordering $\mathcal{T}$, the superoperator notations $o^\times \rho= [o, \rho]$, $o^L \rho= o \rho$, $o^R \rho = \rho o$ for any operators $\rho$ and $o$, $V_S^I(t) = \exp(iH_S^\times t / \hbar) V_S$, and the BCF $L(t)$ defined by
\begin{gather}
    L(t) \equiv \frac{1}{\hbar} {\rm tr}_B [ e^{iH_B^\times t/\hbar} (X_B) X_B \rho_B] \nonumber \\
    = \int_0^\infty \frac{d\omega}{\pi} J(\omega) \left[ \coth \left( \frac{\beta \hbar \omega}{2} \right) \cos(\omega t) - i \sin (\omega t) \right], \label{eq:method_L}
\end{gather}
with $J(\omega) = (\pi/\hbar) \sum_m u_m^2[\delta(\omega-\omega_m) - \delta(\omega+\omega_m)]$ the spectral density.
To model irreversible dynamics, we consider a smooth spectral density $J(\omega)$ representing continuously distributed bath modes.

\subsection{HEOM and pseudomode}
\label{method_HEOM}

To compute \eref{eq:method_rhoS} numerically, we employ HEOM or the pseudomode method.
We begin by deriving the HEOM following \rcite{Ikeda20}.
This involves considering a model BCF $L_{\rm mod}(t)$ expressed with $K$ (generally complex) basis functions $\{ v_k(t) \}_{k=1}^K$ as
\begin{equation}
    L_{\rm mod} (t \geq 0) = \sum_{k=1}^{K} d_k v_k(t) \equiv \bm{d}^\top \bm{v}(t),
    \label{eq:method_Lmod}
\end{equation}
where $d_k \ (k=1,\dots,K)$ are complex numbers and $\top$ denotes the matrix transpose.
We impose the following three assumptions on the set $\{ v_k (t) \}_{k=1}^K$.
First, the vector $\bm{v}(t)$ is closed under time differentiation as $(d/dt) \bm{v}(t) = - Z \bm{v}(t)$ for some $K \times K$ complex matrix $Z$.
Second, the set $\{ v_k (t) \}_{k=1}^K$ is closed under complex conjugation, namely, there exists a complex vector $\bar{\bm{d}} \, ^\top = [\bar{d}_1, \dots, \bar{d}_K]$ such that $(\bm{d}^\top \bm{v}(t))^* = \bar{\bm{d}} \, ^\top \bm{v}(t)$.
Third, there exist a complex vector $\bm{\theta} \in \mathbb{C}^K$ and $K \times K$ complex matrices $D$ and $\bar{D}$ such that $\bm{d}^\top = \bm{\theta}^\top D$, $\bar{\bm{d}} \, ^\top = \bm{\theta}^\top \bar{D}$, and $[D, Z] = [\bar{D}, Z] = 0$.

Under the last two assumptions, the propagator \eref{eq:method_M} with $L_{\rm mod}(t)$ can be expressed as
\begin{equation*}
    \mathcal{M}(t) =
    \exp \left[ - \frac{1}{\hbar} \int_0^t d\tau \,
    V_S^I (\tau)^\times \, \sum_{k=1}^{K} \theta_k \mathcal{Y}_k(\tau) \right],
\end{equation*}
with
\begin{equation*}
\begin{gathered}
    \mathcal{Y}_k(\tau) = \int_0^\tau du \, \\
    \left( [D \, \bm{v}(\tau-u)]_k  V_S^I(u)^L - [\bar{D} \, \bm{v}(\tau-u)]_k V_S^I(u)^R \right).
\end{gathered}
\end{equation*}

To obtain the evolution of $\rho_S(t)$, we calculate the time derivative of $\mathcal{T}[\mathcal{M}(t)]$:
\begin{equation*}
    \frac{d}{dt} \mathcal{T}[\mathcal{M}(t)] = - \frac{1}{\hbar} V_S^I(t)^\times \, \mathcal{T}\left[ \sum_{k=1}^{K} \theta_k \mathcal{Y}_k(t) \mathcal{M}(t) \right].
\end{equation*}
Since the latest time in $\mathcal{T}$ is $t$,  $V_S^I(t)^\times$ can be taken outside the time-ordering. In contrast, $\mathcal{Y}_k(t)$ includes superoperators evaluated at earlier times and must remain inside.
To obtain closed differential equations, we introduce the following auxiliary system operators:
\begin{equation}
\begin{gathered}
    \rho_{\bm{j}} (t) = e^{- i H_S^\times t / \hbar} \, \mathcal{T} \left[
    \prod_{k=1}^{K} \left\{ \frac{ \left( \mathcal{Y}_k (t) \right)^{j_k} }{\sqrt{j_k!}} \right\} \mathcal{M}(t) \right] \rho_S(0),
\end{gathered}
\label{eq:method_rhoj}
\end{equation}
with $\bm{j}^\top = [j_1, \dots, j_{K}]$ and $j_k \in \mathbb{Z}_{\geq 0} \ (k = 1, \dots, K)$.
Note that the reduced density operator corresponds to the element with $\bm{j} = \bm{0}$: $\rho_{\bm{0}} (t) = \rho_S(t)$ [see \eref{eq:method_rhoS}].
These yield HEOM
\begin{equation}
\begin{gathered}
    \frac{d}{dt} \rho_{\bm{j}} (t)
    = - \frac{i}{\hbar}H_S^\times \rho_{\bm{j}} (t) - \sum_{k,k'=1}^{K} Z_{k,k'}^{\bm{j}} \rho_{\bm{j}-\bm{e}_k+\bm{e}_{k'}} (t) \\
    + \sum_{k=1}^{K} \sqrt{j_k} \left( [D \bm{v}(0)]_k V_S^L - [\bar{D} \bm{v}(0)]_k V_S^R \right) \rho_{\bm{j}-\bm{e}_k} (t) \\
    - \frac{1}{\hbar} \sum_{k=1}^{K} \theta_k \sqrt{j_k+1} \, V_S^\times \rho_{\bm{j}+\bm{e}_k},
\end{gathered}
\label{eq:method_HEOM}
\end{equation}
where $\bm{e}_k$ are $K$-dimensional unit vectors with $[\bm{e}_k]_l = \delta_{k,l}$ and $Z_{k,k'}^{\bm{j}} = Z_{k,k'} \sqrt{j_k(j_{k'} + 1 - \delta_{k,k'})}$.

From \eref{eq:method_rhoj}, the initial conditions are $\rho_{\bm{0}}(0) = \rho_S(0)$ and $\rho_{\bm{j} \ne \bm{0}}(0) = 0$.
Solving \eref{eq:method_HEOM} under these conditions yields the reduced density operator from the $\bm{j} = \bm{0}$ element.
Since the HEOMs [\eref{eq:method_HEOM}] are a system of infinitely many coupled differential equations for $\rho_{\bm{j}}(t) \ (\forall\bm{j} \in \mathbb{Z}^K_{\geq 0})$, truncation is necessary for numerical computation.
A common strategy is to limit the hierarchy depth $\sum_{k=1}^{K}j_k$ to a finite cutoff $H$, resulting in $(H+K)!/(H!K!)$ coupled equations for $\rho_{\bm{j}}(t)$ with $\sum_{k=1}^{K}j_k \leq H$ \cite{Tanimura06,SSK07}.
As this number grows rapidly with $K$,
reducing $K$ while preserving the accuracy of $L_{\rm mod}(t)$ is crucial for computational efficiency.

Recent results (\rcites{Link22,Xu23,MT24}) have revealed a close connection between the pseudomode method and the HEOM.
The key difference lies in how the auxiliary states are represented: Although the pseudomode method uses a direct representation, HEOM expresses these states in terms of ladder operator moments.
This representational difference, central to the discussion in \sref{method_moment}, is exemplified in \aref{app:IP}, where we show that a pseudomode equation can be recast into a special case of the HEOM [\eref{eq:method_HEOM}]. Therefore, it suffices to use \eref{eq:method_HEOM} for a unified treatment of both methods.

\subsection{Exact solutions for the harmonic oscillator system}
\label{method_test}

In this section, we focus on the harmonic oscillator system and present practical methods to evaluate exact solutions for a broad class of spectral densities, which are used to benchmark the model BCF.
BCFs defined as \eref{eq:method_L} cannot be exactly represented as \eref{eq:method_Lmod} with finite $K$, implying an inherent error in the model BCF.
To assess its impact on the dynamics, we consider a surrogate harmonic oscillator system [see \eref{eq:method_CT}]:
\begin{equation}
    H_{S, {\rm eff}} = \hbar \omega_0 a^\dagger \! a, \ \ \
    V_S = v_0 \, q,
\label{eq:method_HO}
\end{equation}
with the annihilation (creation) operator $a$ ($a^\dagger$) satisfying $[a,a^\dagger] = 1$, $q = (a+a^\dagger)/\sqrt{2}$, the frequency $\omega_0$, and the coupling parameter $v_0$.

This system is well suited for benchmarking, as the quadratic form of the total Hamiltonian permits exact solutions.
For example, the Heisenberg equations of motion for the system operators can be solved analytically as shown in \aref{exact_EOM}.
The evolution is governed by a kernel function $G_+(t)$ [see \eref{eq:exact_q}], which is the solution of the integro-differential equation
\begin{equation*}
    \frac{d^2}{dt^2} G_+(t) + \frac{\omega_0 v_0^2}{\hbar} \int_0^t d\tau \, \eta(t-\tau) \frac{d}{d\tau}G_+(\tau) + \omega_0^2 G_+ (t) = 0,
\end{equation*}
with initial conditions $G_+(0) = 0$ and $(d/dt) G_+(0) = 1$.
Here, $\eta(t)$ is the friction kernel, defined by
\begin{equation*}
    \eta(t) \equiv 2 \int_0^\infty \frac{d\omega}{\pi} \frac{J(\omega)}{\omega} \cos(\omega t),
\end{equation*}
which relates to the imaginary part of the BCF as $2 {\rm Im}[L(t)] = (d/dt) \eta(t)$, where ${\rm Im}$ denotes the imaginary part.
Beyond solvability, this system exhibits thermalization: If $G_+(t)$ decays sufficiently fast, then $\lim_{t \to \infty} \rho_S(t) = {\rm tr}_B [\exp( - \beta H_{\rm tot})] / {\rm tr} [\exp( - \beta H_{\rm tot})]$ holds, with ${\rm tr}$ the trace over the total space, independent of the initial system state $\rho_S(0)$ \cite{GSI88}.

We denote the equilibrium expectation value of an operator $o$ as $\braket{o}_{\rm eq} = {\rm tr}[o \ \exp( - \beta H_{\rm tot}) ] / {\rm tr} [\exp( - \beta H_{\rm tot})]$.
Building on the non-Markovian tests \cite{Tanimura15,Koyanagi24},
for $o = q$ and $o = p \equiv i(a^\dagger-a)/\sqrt{2}$,
we focus on the equilibrium expectation values of $o^2$, $\braket{o^2}_{\rm eq}$, and the equilibrium autocorrelation functions, $C_{oo}(t) \equiv \braket{\exp(iH_{\rm tot}^\times t/\hbar)(o) \, o}_{\rm eq}$, in the frequency domain
\begin{equation*}
    \mathcal{F}[C_{oo}] (\omega) \equiv \int_{-\infty}^\infty dt \, C_{oo}(t) e^{i \omega t}.
\end{equation*}
Note that the Fourier transform can be calculated as $\mathcal{F}[C_{oo}] (\omega) = 2 {\rm Re}[ \int_{0}^\infty dt \, C_{oo}(t) \exp(i \omega t)]$ due to $C_{oo}(-t) = [C_{oo}(t)]^*$ and satisfies the fluctuation-dissipation relation $\mathcal{F}[C_{oo}] (- \omega) = e^{-\beta \hbar \omega} \mathcal{F}[C_{oo}](\omega)$ \cite{Zhang23}.
Although the original tests included computing higher-order correlation functions \cite{Tanimura15}, we restrict our analysis to second-order ones, as the Gaussianity ensures that higher-order correlations are entirely determined by them.

Analytic expressions for these quantities are available \cite{GSI88} and are presented explicitly in \aref{app:exact_solutions}.
A key observation is that their evaluation requires only the Laplace transform of the friction kernel,
\begin{equation*}
    \hat{\eta}(s) = \int_0^\infty dt \, \eta(t) \, e^{-s t} \ ({\rm Re}(s) \geq 0),
\end{equation*}
where ${\rm Re}$ denotes the real part.
Inserting the definition of $\eta(t)$, $\hat{\eta}(s)$ can be expressed with $J(\omega)$ as
\begin{equation}
    \hat{\eta}(s) = \frac{2 s}{\pi} \int_0^\infty d\omega' \frac{J(\omega')/\omega'}{(\omega')^2 + s^2}.
    \label{eq:method_etas}
\end{equation}
Specifically, we require $\hat{\eta}(|s|) \ (|s| > 0)$ for computing $\braket{o^2}_{\rm eq}$ and $\hat{\eta}(-i\omega) \ (\omega \in \mathbb{R})$ for $\mathcal{F}[C_{oo}] (\omega)$ ($o=q,p$).

In \aref{app:exact_analytic_eta}, we present analytic expressions for $\hat{\eta}(s)$ in several cases.
When such expressions are unavailable, numerical evaluation is required.
For $\hat{\eta}(|s|) \ (|s| > 0)$, the integral in \eref{eq:method_etas} can be computed directly.
For $\hat{\eta}(- i\omega) \ (\omega \in \mathbb{R})$,
on the other hand, the integrand has a pole at $\omega' = |\omega|$.
By deforming the contour to avoid the pole, we obtain
\begin{equation}
    \hat{\eta}(- i \omega) = \frac{J(\omega)}{\omega} - i \frac{2 \omega}{\pi} \, {\rm p.v.} \! \int_0^\infty d\omega' \frac{J(\omega')/\omega'}{(\omega')^2 - \omega^2},
    \label{eq:method_eta-iomega}
\end{equation}
where ${\rm p.v.}$ denotes the Cauchy principal value.
This expression is amenable to numerical evaluation and also provides analytic insight into the behavior of $\mathcal{F}[C_{qq}](\omega)$ near $\omega = 0$ [see \eref{eq:exact_FCqq_near_0}].

An alternative way to evaluate $\hat{\eta}(s)$ is to fit the spectral density $J(\omega)$ using simple functions.
A promising approach is to approximate ${\rm Im}[L(t)]$ or $\eta(t)$ by a finite (though possibly large) sum of complex exponentials:
\begin{equation}
  \begin{gathered}
    {\rm Im}[L(t \geq 0)] = - 2 \sum_j {\rm Im} \left[ c_j e^{- \mu_j t} \right] \ \ \ {\rm or} \\
    \eta(t \geq 0) = 4 \sum_j {\rm Im} \left[ (c_j / \mu_j) e^{- \mu_j t} \right],
  \end{gathered}
  \label{eq:method_GMT}
\end{equation}
where $\{c_j, \mu_j\}_j$ are complex fitting parameters with ${\rm Re} (\mu_j) > 0$ for the stability.
The corresponding spectral density $J(\omega)$ follows from the Fourier transform [see \eref{eq:GMT_J}].
If ${\rm Im}(c_j) = 0$ is assumed, $J(\omega)$ is given by
\begin{equation}
    J(\omega) = \sum_j \frac{p_j \omega}{\left[ (\omega+\Omega_j)^2+\Gamma_j^2 \right] \left[ (\omega-\Omega_j)^2+\Gamma_j^2 \right]},
    \label{eq:method_MT}
\end{equation}
with $\Gamma_j = {\rm Re}(\mu_j)$, $\Omega_j = {\rm Im}(\mu_j)$, and $p_j = 8 {\rm Re}(c_j) \Omega_j \Gamma_j$.
This decomposition was originally proposed by Meier and Tannor \cite{MT99}, and \eref{eq:method_GMT} serves as its generalization.
In \aref{app:GMT}, we explore this model in more detail, including a fitting procedure.
Given $\{c_j, \mu_j \}_j$, the analytic expression for $\hat{\eta}(s)$ valid for ${\rm Re}(s) \geq 0$ is available [see \eref{eq:exact_eta_GMT}], enabling evaluation of the exact solutions.

\subsection{Moment representation of the oscillator state}
\label{method_moment}

In this section, we introduce a representation of the oscillator state that enables a reliable and computationally efficient testing of a model BCF.
Previous non-Markovian tests \cite{Tanimura15,Koyanagi24} solved the quantum Fokker-Planck equation, which is the Wigner representation of the HEOM [\eref{eq:method_HEOM}]. Alternatively, the conventional Fock basis can be used to represent the oscillator system state.
Since the oscillator Hilbert space is infinite-dimensional, truncation is necessary for numerical computations.
Under strong system-bath coupling, both the effective system dimension $n_S$ and the required hierarchy depth $H$ increase, limiting the applicability of these approaches to simple cases with small $K$ [the number of terms in $L_{\rm mod}(t)$].
Additionally, truncating the oscillator state introduces another source of error, obscuring whether deviations in the results stem from $L_{\rm mod}(t)$ or the state truncation.

To address these issues, we adopt the moment representation of the oscillator state:
\begin{equation}
  \phi_{m,n,\bm{j}} (t) \equiv \frac{{\rm tr}_S(a^m \rho_{\bm{j}} (t) (a^\dagger)^n )}{\sqrt{m! \, n!}},
  \label{eq:method_moment}
\end{equation}
where ${\rm tr}_S$ denotes the trace over the system.
For $\bm{j} = 0$, the set $\{ \phi_{m,n,\bm{0}} (t) \}_{m,n = 0}^\infty$ corresponds to the moments of the ladder operators.
A detailed discussion of this representation is provided in \aref{app:moment}.
In \aref{app:moment_HEOM}, we derive the evolution equation for $\{ \phi_{m,n,\bm{j}} (t) \}$, namely, the HEOM [\eref{eq:method_HEOM}] in this representation.
This reveals that the evolution of each element $\phi_{m,n,\bm{j}} (t)$, with depth $\mathcal{H} \equiv m + n + \sum_{k=1}^K j_k$, depends only on elements with depth $\mathcal{H}$ or $\mathcal{H}-2$.
This decoupling follows from the use of the moment representation for the auxiliary states in HEOM, as discussed at the end of \sref{method_HEOM}, and the Gaussian nature of the total system.

This decoupling enables an exact truncation in the moment representation.
For example, $\braket{o^2}_{\rm eq}$ and $\mathcal{F}[C_{oo}](\omega)$ for $o=q,p$ are second-order moments, and their computation only requires elements $\{ \phi_{m,n,\bm{j}} \}$ with $m + n + \sum_{k=1}^K j_k \leq 2$.
This set contains merely $(K+4)!/(2!(K+2)!)$ complex numbers, in sharp contrast with the ordinary HEOM containing $n_S^2 (H+K)!/(H!K!)$ complex numbers.
The reduction enables efficient computation of the dynamics even for large $K$.
Moreover, since the truncation is exact, we can isolate and assess the impact of the approximation in $L_{\rm mod}(t)$, enabling reliable testing.

\section{Accuracy analysis}
\label{analysis}

Before proceeding to the testing procedure, here we examine recently proposed fitting methods for constructing a model BCF.
Using the exact methodologies developed in the previous section, we assess how well these fitting methods can describe thermalization.

The fitting methods assume that the BCF is expressed as a linear combination of exponential functions,
\begin{equation}
    L_{\rm mod}(t \geq 0) = \sum_{k=1}^K d_k e^{-z_k t},
    \label{eq:analysis_Lmod_exp}
\end{equation}
where $z_k \in \mathbb{C}$ with ${\rm Re}(z_k) > 0$ for stability.
This corresponds to $v_k(t) = \exp(-z_k t)$ in \eref{eq:method_Lmod}, and the assumptions in \sref{method_HEOM} are satisfied by taking the set $\{z_k\}_{k=1}^K$ closed under complex conjugation and setting $\bm{\theta} = [1 \ 1 \cdots 1]^\top$, $D = {\rm diag}[d_1 \ d_2 \cdots d_K]$, and $\bar{D} = {\rm diag}[\bar{d}_1 \ \bar{d}_2 \cdots \bar{d}_K]$.
The resulting HEOM [\eref{eq:method_HEOM}] reads
\begin{equation}
\begin{gathered}
    \frac{d}{dt} \rho_{\bm{j}} (t)
    = - \left[ \frac{i}{\hbar}H_S^\times + \sum_{k=1}^K z_k j_k \right] \rho_{\bm{j}} (t) \\
    + \sum_{k=1}^{K} \sqrt{j_k} \left( d_k V_S^L - \bar{d}_k V_S^R \right) \rho_{\bm{j}-\bm{e}_k} (t) \\
    - \frac{1}{\hbar} \sum_{k=1}^{K} \sqrt{j_k+1} \, V_S^\times \rho_{\bm{j}+\bm{e}_k}.
\end{gathered}
\label{eq:analysis_HEOM}
\end{equation}

We summarize the fitting methods in \sref{analysis_fit}.
In \sref{analysis_error}, we compare their performance for a featureless Ohmic spectral density.
Lastly, in \sref{analysis_sub}, we examine a challenging scenario of a finite-temperature bath with a sub-Ohmic spectral density.

\subsection{Fitting methods}
\label{analysis_fit}

Here, we introduce four methods for fitting the BCF in the form of \eref{eq:analysis_Lmod_exp}.
The following descriptions focus on a broad overview of each method, and technical details are deferred to the cited references and Appendixes.

\begin{enumerate}[label=(\roman*)]
    \item {\it Adaptive Antoulas-Anderson (AAA) algorithm}.
    The algorithm offers a rational approximation for real or complex functions \cite{NST18}, and was applied in \rcite{Xu22} to fit the Fourier transform of the BCF:
    \begin{equation}
        \mathcal{F}[L](\omega) = \frac{2 J (\omega)}{1-e^{-\beta \hbar \omega}}.
        \label{eq:analysis_Lomega}
    \end{equation}
    It models the BCF in the barycentric form as
    \begin{equation*}
        \mathcal{F}[L_{\rm mod}](\omega) = \frac{\sum_j \mathcal{F}[L](\omega_j) W_j / (\omega - \omega_j)}{ \sum_j W_j/(\omega-\omega_j)},
    \end{equation*}
    with the parameters $\{ \omega_j, W_j \}_j$ determined iteratively.
    A partial fraction decomposition then expresses $\mathcal{F}[L_{\rm mod}](\omega)$ as a sum over simple poles, which yields the time-domain expression as \eref{eq:analysis_Lmod_exp}.

    \item {\it Interacting pseudomode (IP) approach.}
    In \rcites{Medina21,Lednev24}, the authors fit the Fourier transform of the BCF using the ansatz,
    \begin{equation}
        \mathcal{F}[L_{\rm mod}](\omega) = \frac{1}{\pi} \bm{l}^\top {\rm Im} \left[ (\tilde{\omega} - \omega)^{-1} \right] \bm{l},
        \label{eq:analysis_IPansatz}
    \end{equation}
    where the fitting parameters are a real vector $\bm{l}$ and a complex matrix $\tilde{\omega}$ with elements $\tilde{\omega}_{k,k'} = \omega_{k,k'} - (i/2) \delta_{k,k'} \kappa_k$ with a real symmetric $\omega_{k,k'}$ and positive $\kappa_k$.

    This form arises from an interacting pseudomode model, as elaborated in \aref{app:IP}.
    Note that we have $\mathcal{F}[L_{\rm mod}](\omega) > 0$ for any $\omega$, which ensures complete positivity of the reduced dynamics \cite{Diosi14}.
    Diagonalizing $\tilde{\omega}$ expresses $\mathcal{F}[L_{\rm mod}](\omega)$ as a sum over simple poles, leading to the time-domain expression as \eref{eq:analysis_Lmod_exp}.
    We note that this approach has also gained attention for its potential application in quantum algorithms for simulating open quantum dynamics \cite{Lambert24,Sun24}.

    \item {\it Estimation of Signal Parameters via Rotational Invariance Techniques (ESPRIT).}
    For equidistant data points $\{ L(t_n) \}_n$ with $t_n = n \Delta t_{\rm data}$, ESPRIT estimates $\{ d_k, z_k \}_{k=1}^K$ by minimizing $\sum_n | L(t_n) - L_{\rm mod}(t_n) |$, using the translational identity $\exp(-z_k t_m) \exp(-z_k t_n) = \exp(-z_k t_{m+n})$ \cite{PT13}.
    Its application to BCF fitting was proposed in \rcite{Takahashi24}, demonstrating superior performance over AAA and other time-domain fitting methods.

    \item {\it Generalized Meier-Tannor with FITted Matsubara modes (GMT$\&$FIT).}
    In \rcite{Brieuc24}, a two-step approach was proposed.
    First, the spectral density is fitted using the Maier-Tannor form [\eref{eq:method_MT}].
    The real part of the BCF ${\rm Re}[L(t)]$ involves an infinite sum of real exponentials due to Matsubara contributions.
    Second, this sum is approximated by a finite set of real exponentials, as originally done in \rcite{Lambert19}.

    We propose a modification to the first step by adopting the more general form [\eref{eq:method_GMT}].
    The resulting ${\rm Re}[L(t)]$ again contains an infinite sum [see \eref{eq:GMT_ReL}], which we treat using the same second step. Further methodological details are provided in \aref{app:GMT}.

\end{enumerate}

\begin{table}[h]
  \centering
  \begin{tabular}{|c|c|c|} \hline
    & Linear & Nonlinear \\ \hline
    Time & ESPRIT & GMT$\&$FIT \\ \hline
    Fourier & AAA & IP \\ \hline
  \end{tabular}
  \caption{
  Classification of the four fitting methods.
  }
  \label{tab:analysis_4alg}
\end{table}

Methodologically, these can be classified as shown in \tref{tab:analysis_4alg}.
ESPRIT and GMT$\&$FIT operate in the time domain, while AAA and IP work in the frequency domain.
ESPRIT and AAA avoid nonlinear parameter dependence by reducing the problem to linear optimization, whereas GMT$\&$FIT and IP rely on nonlinear optimization methods.

Previous assessments of these methods have primarily focused on the reproducibility of the BCF or qualitative features of the dynamics.
While quantitative consistency with other methods has been confirmed in \rcites{Xu22,Lednev24}, these studies mainly addressed the transient regime.
Here, using a harmonic oscillator system, we assess the accuracy of thermalization simulations by direct comparison with the exact solutions.

\begin{figure}[t]
  \includegraphics[keepaspectratio, scale=0.38]{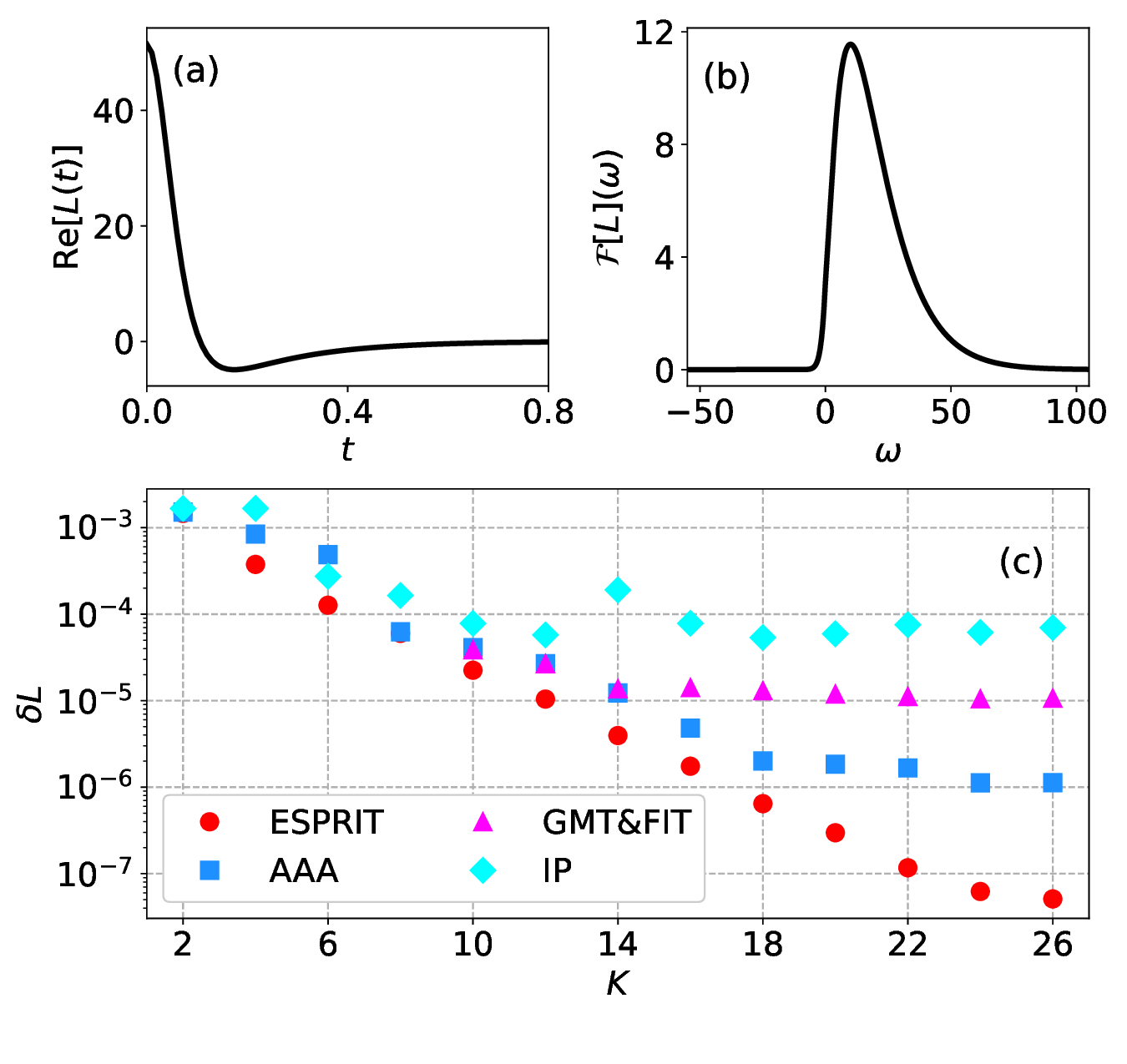}
  \caption{
  Error in the model BCF from the four fitting methods.
  Exact BCF in the time (a) and frequency (b) domains. All quantities are shown in units of $\hbar = \omega_0 = v_0 = 1$.
  (c) Error $\delta L$ defined by \eref{eq:analysis_dL} as a function of $K$.
  }
  \label{fig:analysis_ohmic_L}
\end{figure}

\subsection{Comparison for an Ohmic spectral density}
\label{analysis_error}

In this section, we apply the four fitting methods to an Ohmic spectral density and compare their performance in describing thermalization. Among them, ESPRIT achieves the smallest error for a fixed $K$.

We consider spectral densities with an exponential high-frequency cutoff,
\begin{equation}
  J_{\rm exp}(\omega \geq 0) = \frac{\pi}{2} \alpha \omega_c^{1-s} \omega^s e^{- \omega / \omega_c},
  \label{eq:analysis_Jexp}
\end{equation}
where $\alpha$ is the coupling strength and $\omega_c$ is the cutoff frequency. The exponent $s$ (Ohmicity) characterizes the low-frequency behavior.
In this section, we focus on the Ohmic case ($s = 1$),
for which the Laplace transform of the friction kernel $\eta(t)$, required for the exact solutions (\sref{method_test}), can be evaluated accurately [see \erefs{eq:exact_eta_exp_positive} and (\ref{eq:exact_eta_exp_imaginary})].
This case therefore permits a stringent test of the BCF.
In units of $\hbar = \omega_0 = v_0 = 1$, we arbitrarily set $\alpha = 1$, $\omega_c = 10$, and $\beta = 1$.
The corresponding BCFs are shown in \frefs{fig:analysis_ohmic_L}(a) and \ref{fig:analysis_ohmic_L}(b).

Model BCFs are constructed using the four fitting methods introduced in \sref{analysis_fit}.
The data selection procedure is described in \aref{app:num_analysis_error}.
We quantify the relative error in the model BCFs as
\begin{equation}
  \delta L \equiv \frac{1}{t_f} \int_0^{t_f} dt \, \left| \frac{L(t) - L_{\rm mod}(t)}{L(0)} \right|.
  \label{eq:analysis_dL}
\end{equation}
Since our main interest lies in simulating thermalization, we set $t_f = 30$, the time required to reach a steady state.
A relative error of $\delta L \lesssim 10^{-4}$ ensures nearly perfect agreement between $L(t)$ and $L_{\rm mod}(t)$ in both time and frequency domains on the scales of \frefs{fig:analysis_ohmic_L}(a) and \ref{fig:analysis_ohmic_L}(b).

The dependence of $\delta L$ on $K$ is shown in \fref{fig:analysis_ohmic_L}(c).
We observe that ESPRIT consistently achieves the lowest error across all values of $K$.
AAA and GMT$\&$FIT provide indistinguishable results for $K = 10,12,$ and $14$.
For $K \geq 16$, GMT$\&$FIT plateaus, while AAA continues to improve.
A distinctive feature of GMT$\&$FIT is its use of real exponential functions, enabling a broader mix of decay rates than fitting with complex exponential functions at fixed $K$. However, this result indicates that such mixing does not significantly enhance the fitting accuracy.
Lastly, we observe that IP gives the largest error among the methods considered here.
This implies that the positivity constraint $\mathcal{F}[L_{\rm mod}](\omega) > 0 \, (\forall \omega)$ may be overly restrictive at the ansatz level.

\begin{figure}[t]
  \includegraphics[keepaspectratio, scale=0.4]{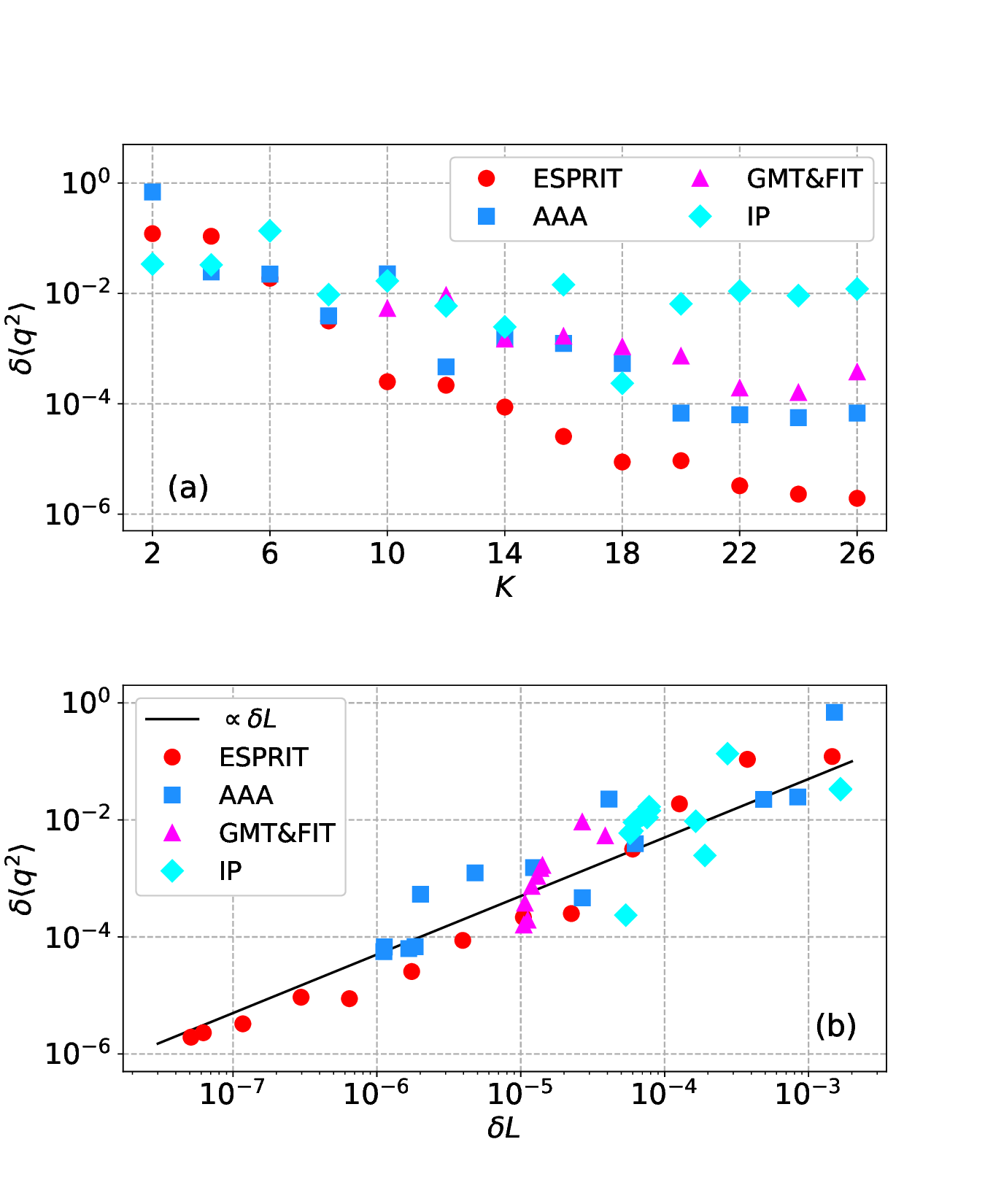}
  \caption{
  Error in the steady-state expectation value, $\delta \! \braket{q^2}$, defined by \eref{eq:analysis_do2}, as a function of (a) $K$ and (b) $\delta L$.
  In panel (b), the solid black line indicates the linear relation $\delta \! \braket{q^2} = 50 \, \delta L$.
  }
  \label{fig:analysis_ohmic_q2error}
\end{figure}

Using the model BCFs, we compute the reduced dynamics $\rho_S(t)$ by solving the HEOM [\eref{eq:analysis_HEOM}] (see \aref{app:num_analysis_error} for computational details).
The steady-state expectation values and the autocorrelation functions are evaluated as
\begin{equation}
    \braket{o^2}_{\rm mod} \equiv {\rm tr}_S [o^2 \rho_S (t_f)],
    \label{eq:analysis_omod}
\end{equation}
and
\begin{equation}
    C^{\rm mod}_{oo}(t) \equiv \braket{\exp(iH_{\rm tot}^\times t/\hbar)(o) \, o}_{\rm mod},
    \label{eq:analysis_Coomod}
\end{equation}
for $o=q,p$, where the subscript ${\rm mod}$ indicates evaluation using a model BCF.
As the results for $o = q$ and $p$ show similar behavior, we present only the case $o = q$ below.

Figure~\ref{fig:analysis_ohmic_q2error}(a) shows the $K$ dependence of the relative error in the expectation value, defined as
\begin{equation}
    \delta \! \braket{o^2} \equiv \frac{|\braket{o^2}_{\rm eq} - \braket{o^2}_{\rm mod}|}{\braket{o^2}_{\rm eq}}.
    \label{eq:analysis_do2}
\end{equation}
We observe that it resembles the behavior in \fref{fig:analysis_ohmic_L}(c), the error in the model BCF.
Previous work \rcite{Huang24} showed that the error in expectation values can be bounded from above by $\delta L$.
While that analysis assumed a finite-dimensional system, it is worthwhile to examine the connection in the current setting.
To this end, \fref{fig:analysis_ohmic_q2error}(b) shows $\delta \! \braket{q^2}$ plotted against $\delta L$.
We find that the trend can be well captured by the linear relation, $\delta \! \braket{q^2} \propto \delta L$, as indicated by the solid black line in the figure.
This scaling is consistent with the analytic bound in \rcite{Huang24} in the small-$\delta L$ limit.
However, we note that the bound grows exponentially with time and does not tightly constrain the long-time (steady-state) error.

\begin{figure}[t]
  \includegraphics[keepaspectratio, scale=0.34]{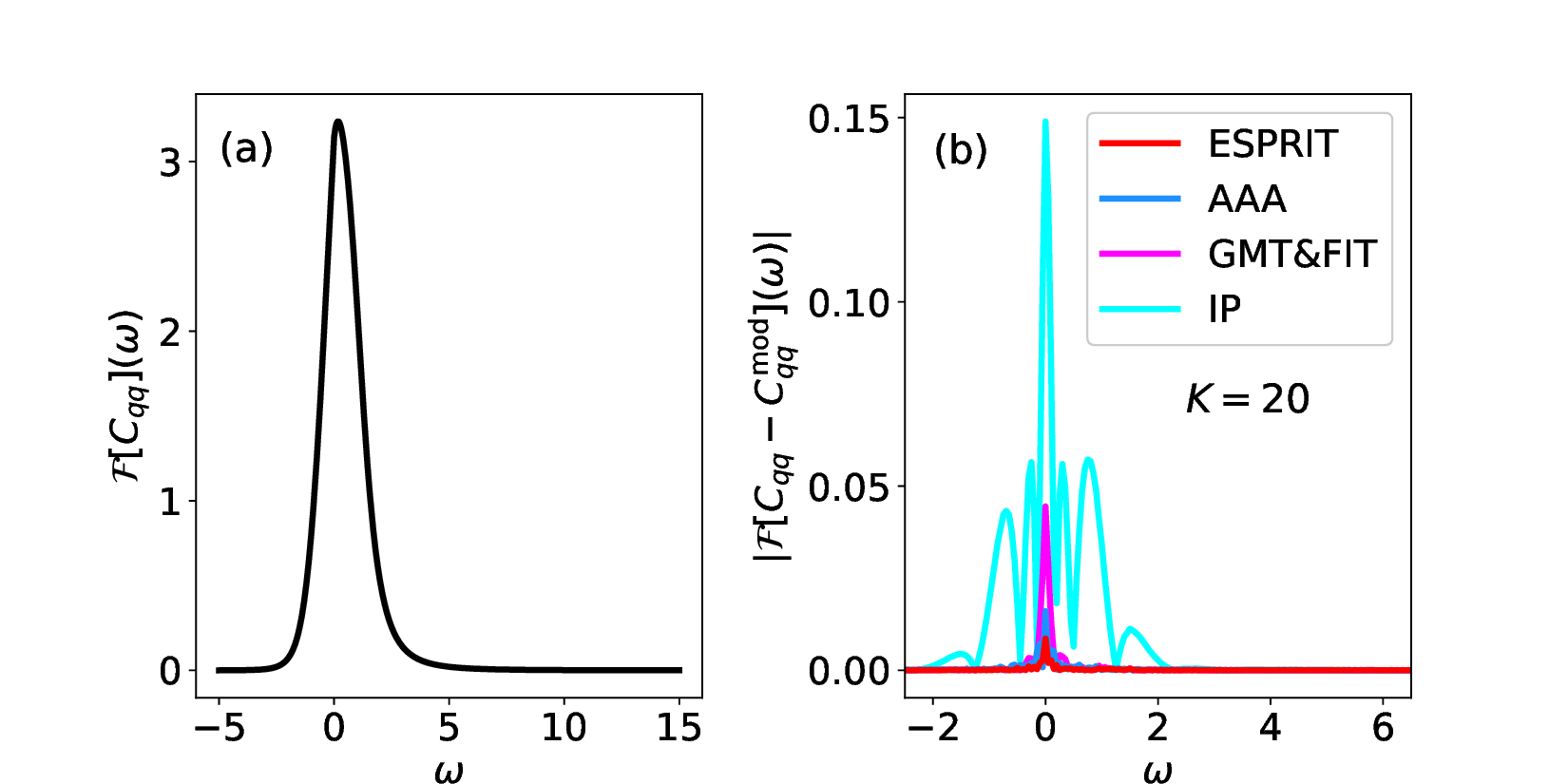}
  \caption{
  Error in the autocorrelation function.
  All dimensional quantities are shown in units of $\omega_0 = 1$.
  (a) Exact equilibrium autocorrelation function.
  (b) Deviation of the Fourier transform for the four fitting methods with $K = 20$.
  }
  \label{fig:analysis_ohmic_corr_fourier}
\end{figure}

We next analyze the Fourier transform of the autocorrelation function.
Figure~\ref{fig:analysis_ohmic_corr_fourier}(a) shows the exact result.
On this scale, the ESPRIT results for $K \geq 8$ are indistinguishable from the exact results, while the IP result shows noticeable deviation near $\omega = 0$ even for $K = 16$ (not shown).
To highlight the difference, \fref{fig:analysis_ohmic_corr_fourier}(b) plots the deviation $|\mathcal{F}[C_{qq} - C^{\rm mod}_{qq}](\omega)|$.
Overall, ESPRIT gives the smallest error, whereas IP yields the largest, consistent with the earlier observations on the fitting accuracy.

The accuracy of a model BCF depends on the dataset used for fitting.
Here, the dataset is generated by equidistant sampling in both time and frequency domains for a featureless Ohmic spectral density (\aref{app:num_analysis_error}).
Further improvement may be achieved by using a more carefully calibrated dataset.

\subsection{Sub-Ohmic case}
\label{analysis_sub}

As a challenging case, we examine a finite-temperature bath with a sub-Ohmic spectral density. While ESPRIT performed best in \sref{analysis_error}, this study reveals its limitations for systems with long-time tails in $L(t)$.

\begin{figure}[t]
  \includegraphics[keepaspectratio, scale=0.38]{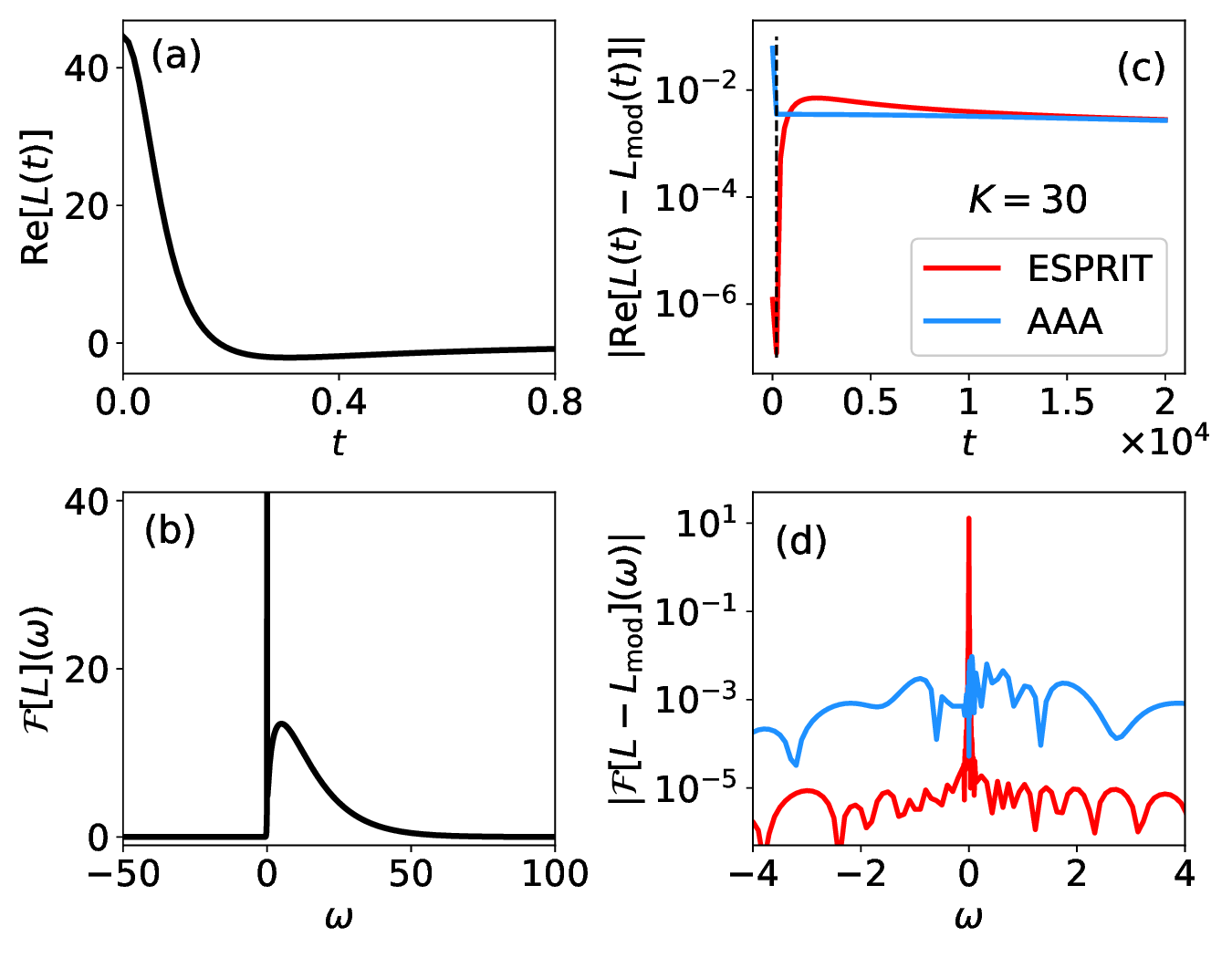}
  \caption{
  Error in the model BCF from ESPRIT and AAA with $K = 30$.
  All quantities are shown in units of $\hbar = \omega_0 = v_0 = 1$.
  (a), (b) Exact BCF in the (a) time and (b) frequency domains.
  (c), (d) Error in (c) ${\rm Re}[L_{\rm mod}(t)]$ and (d) $\mathcal{F}[L_{\rm mod}](\omega)$ for ESPRIT (red curve) and AAA (blue curve).
  In panel (c), the thin dashed vertical black line shows $t = 200$, the fitting range for ESPRIT.
  In panel (d), the region $|\omega| < 4 \times 10^{-4}$, where $\mathcal{F}[L](\omega) > 50$ is excluded.
  }
  \label{fig:analysis_subohmic_L}
\end{figure}

We consider \eref{eq:analysis_Jexp} with parameters $s = 0.5$, $\alpha = 1$, $\omega_c = 10$, and $\beta = 10$ in units of $\hbar = \omega_0 = v_0 = 1$.
The resulting BCFs are shown in \frefs{fig:analysis_subohmic_L}(a) and \ref{fig:analysis_subohmic_L}(b).
At finite temperatures, $\mathcal{F}[L](\omega) \simeq 2 J(\omega) / (\beta \hbar \omega)$ near $\omega = 0$ [see \eref{eq:analysis_Lomega}], leading to a divergence at $\omega = 0$ in sub-Ohmic cases, as seen in \fref{fig:analysis_subohmic_L}(b).
This divergence corresponds to a long tail in ${\rm Re} [L(t)]$, posing a previously unrecognized challenge absent in the Ohmic case.

Here, we examine ESPRIT and AAA.
For ESPRIT, data points are equidistantly sampled up to $t_{\rm max} = 200$.
For AAA, which operates in the frequency domain, the regions near $\omega = 0$ and at large $|\omega|$ are treated separately to capture the divergent and smooth behaviors of $\mathcal{F}[L](\omega)$, respectively.
Further details are given in \aref{app:num_analysis_sub}.

On the scales of \frefs{fig:analysis_subohmic_L}(a) and \ref{fig:analysis_subohmic_L}(b), we observe that the ESPRIT and AAA results with $K = 30$ are indistinguishable from the exact curves (not shown).
In \rcite{Takahashi24}, AAA was applied to a sub-Ohmic spectral density, and noticeable deviations in ${\rm Re}[L(t)]$ were reported even at short times.
The present agreement indicates that the discrepancy is not inherent to AAA, but likely arises from inadequate data sampling, which in \rcite{Takahashi24} was logarithmic over the full frequency range.

Figure~\ref{fig:analysis_subohmic_L}(c) shows the error in ${\rm Re}[L_{\rm mod}(t)]$ in the long-time region.
The thin dashed vertical line indicates $t = t_{\rm max}$.
ESPRIT yields significantly smaller error for $t \leq t_{\rm max}$, where the fitting is performed.
However, the ESPRIT error grows abruptly for $t > t_{\rm max}$ and exceeds that of AAA for $t \geq 800$.
Figure~\ref{fig:analysis_subohmic_L}(d) shows the error in $\mathcal{F}[L_{\rm mod}](\omega)$ in the vicinity of $\omega = 0$ (excluding $|\omega| < 4 \times 10^{-4}$).
While ESPRIT achieves smaller overall error, it displays a pronounced peak for $|\omega| < 2 \times 10^{-2}$, where AAA yields higher accuracy,
as expected from its fitting range that includes this low-frequency region.

\begin{figure}[t]
  \includegraphics[keepaspectratio, scale=0.45]{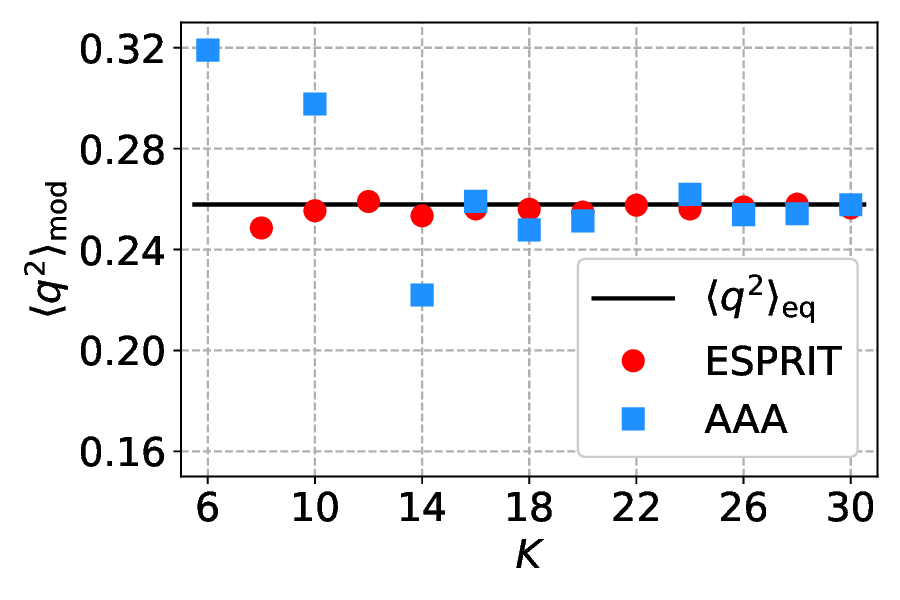}
  \caption{
  Comparison of the exact equilibrium expectation value (black line) with the steady-state values obtained with ESPRIT (red circles) and AAA (blue squares) as a function of $K$.
  }
  \label{fig:analysis_subohmic_q2}
\end{figure}

Now, we assess how errors in the model BCFs impact thermalization simulations.
Computational details are given in \aref{app:num_analysis_sub}.
For $o = p$, the analysis is analogous to the Ohmic case discussed above, so we focus on $o = q$ in the following.

\begin{figure}[t]
  \includegraphics[keepaspectratio, scale=0.32]{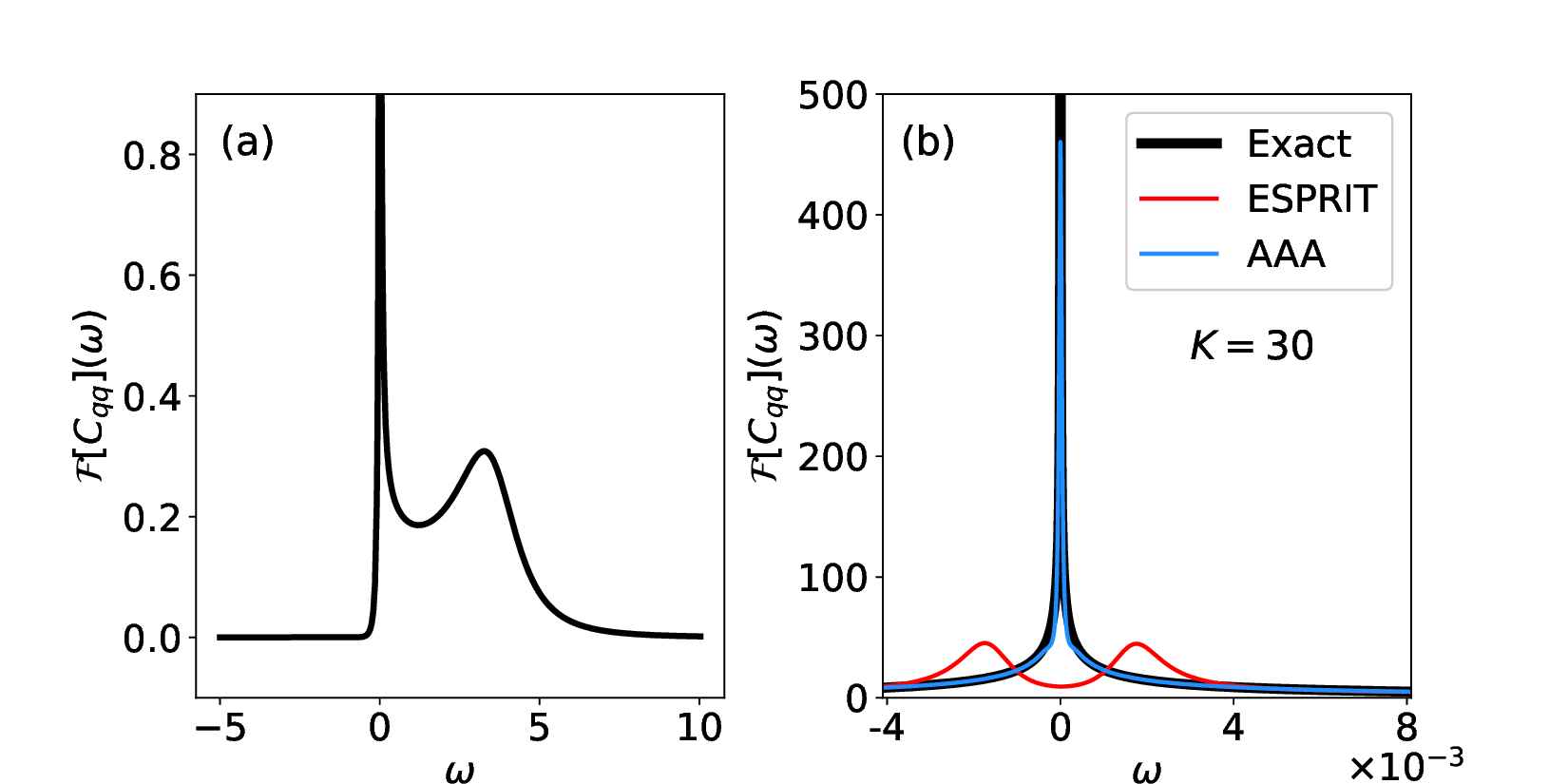}
  \caption{
  Error in the autocorrelation function.
  All quantities are shown in units of $\omega_0 = 1$.
  (a) Exact equilibrium autocorrelation function.
  (b) Magnified view around $\omega = 0$, comparing the exact result (thick black) with ESPRIT (thin red) and AAA (thin blue) results for $K = 30$.
  }
  \label{fig:analysis_subohmic_corr_fourier}
\end{figure}

Figure~\ref{fig:analysis_subohmic_q2} compares the exact equilibrium expectation value $\braket{q^2}_{\rm eq}$ with $\braket{q^2}_{\rm mod}$ for various $K$.
We see that the ESPRIT results are notably more stable and accurate for small $K$, likely due to the smaller error in the BCF over the region $t \leq t_{\rm max}$.
Figure~\ref{fig:analysis_subohmic_corr_fourier}(a) shows the exact autocorrelation function $\mathcal{F}[C_{qq}](\omega)$.
On this scale, both AAA and ESPRIT with $K = 30$ reproduce the exact curves, including the divergence near $\omega = 0$ (not shown).
Such divergence is characteristic of sub-Ohmic spectral densities [see \eref{eq:exact_FCqq_near_0}].

To examine the reproducibility of the divergence more closely, \fref{fig:analysis_subohmic_corr_fourier}(b) shows a zoom around $\omega = 0$.
We observe that AAA captures the divergent behavior well up to $\mathcal{F}[C_{qq}](\omega) \simeq 450$, whereas ESPRIT fails in the region $\omega \in [-4 \times 10^{-3}, 4 \times 10^{-3}]$.
We confirm that increasing $K$ in ESPRIT yields only marginal improvement.
This failure stems from the poor reproduction of ${\rm Re}[L(t)]$ in the long-time region discussed above.
As detailed in \aref{app:subESPRIT}, extending the fitting range $t_{\rm max}$ improves the accuracy of ${\rm Re}[L_{\rm mod}(t)]$ at long times and yields a better description of the divergent behavior.
However, since ESPRIT requires equidistant data points together with small grid spacings to resolve short-time features, $t_{\rm max}$ is practically limited, thereby constraining the accuracy achievable with ESPRIT.

Exact analysis of a dephasing model revealed nonexponential long-time decoherence \cite{Trushechkin24}.
It was argued from this result that methods based on autonomous linear systems (e.g., HEOM and the pseudomode method), which at most yield polynomial-exponential time dependence, can only approximate dynamics over finite intervals.
The divergence of $\lim_{\omega \to 0} \mathcal{F}[C_{qq}](\omega)$ signals the non-exponential decay of $C_{qq}(t)$.
Nevertheless, excellent agreement with AAA is observed in \fref{fig:analysis_subohmic_corr_fourier}(b), motivating further investigation of complex spectral densities using these methods.

\section{Demonstration}
\label{demo}

In this section, we present our method for testing a model BCF using surrogate harmonic oscillator systems [\eref{eq:method_HO}].
Section~\ref{demo_test} outlines the testing procedure, and \sref{demo_example} illustrates its application through several examples.

\subsection{Testing procedure}
\label{demo_test}

In this section, we develop the testing procedure based on a perturbative master equation, where the nonunitary part is characterized by system state transitions induced by $V_S$.
By assigning a surrogate harmonic oscillator to each transition, we estimate error in the target dynamics.

Given the total Hamiltonian [\eref{eq:method_Htot}] with a BCF $L(t)$, our goal is to accurately simulate thermalization, namely, relaxation of the system state to the reduced Gibbs state, independent of its initial state, which we assume for the present system.
We simulate the dynamics using HEOM or the pseudomode method, both relying on a model BCF $L_{\rm mod}(t)$
[\eref{eq:method_Lmod}].
Inaccuracies of $L_{\rm mod}(t)$ are a primary source of error in these methods, and to assess its validity we propose using surrogate harmonic oscillator systems [\eref{eq:method_HO}] with the same BCF.

We focus on equilibrium expectation values and equilibrium correlation functions, which are central to studies of thermalization \cite{SHT18,Tanimura20,LMR23,Jankovic23}.
Because these quantities can be sensitive to different regions of the BCF, we assess them separately: The accuracy of the equilibrium expectation values (correlation functions) is tested by checking the reproducibility of $\braket{o^2}_{\rm eq}$ ($C_{oo}(t)$) for $o = q, p$ in a surrogate harmonic oscillator system.
In this system, higher-order correlations are determined by second-order ones, so it suffices to examine $\braket{o^2}_{\rm eq}$ or $C_{oo}(t)$ ($o = q, p$) to evaluate the accuracy of each quantity \cite{cpq}.

The harmonic oscillator system contains two free parameters: $\omega_0$ and $v_0$.
Even for a fixed $L_{\rm mod}(t)$, the error generally depends on their values (see \sref{demo_example}),
so they should be chosen to reflect the characteristics of the present system.
To this end, we refer to the Redfield equation \cite{Breuer02,Redfield57}, a perturbative master equation whose nonunitary part is characterized by
\begin{equation*}
    \int_0^\infty dt \, L(t) \, e^{-iH_S^\times t / \hbar} (V_S) = \sum_{\Omega \geq 0} \left(\Gamma(\Omega) C_\Omega + \Gamma(-\Omega) C_\Omega^\dagger \right)
\end{equation*}
where $C_\Omega = \sum_{i \leq j} \delta_{\Omega_j - \Omega_i, \Omega} \left(1 - \delta_{i,j}/2 \right) \ketbra{i}{i} \! V_S \! \ketbra{j}{j}$ with $H_S \ket{i} = \hbar\Omega_i \ket{i} \ (i \leq j \Rightarrow\Omega_i \leq\Omega_j)$ and $\Gamma(\Omega) = \int_0^\infty dt \, L(t) \, \exp(i \Omega t)$.
We note $V_S = \sum_{\Omega \geq 0} (C_\Omega + C_{\Omega}^\dagger)$.
In the harmonic oscillator case [\eref{eq:method_HO}], only a single Bohr frequency contributes as $C_\Omega^{\rm HO} = \delta_{\Omega,\Omega_0} v_0 e^{-z_0}/\sqrt{2} \, ( \cosh(z_0) a + \sinh(z_0) a^\dagger)$ with $\Omega_0 = \omega_0 \sqrt{1 + 2 \lambda v_0^2 / (\hbar \omega_0)}$ and $z_0 = \tanh^{-1}(\lambda v_0^2 / (\hbar \omega_0 + \lambda v_0^2)) / 2$.
For a given Bohr frequency $\Omega$, the BCF enters only through $\Gamma(\Omega)$.
We therefore require $\Omega = \Omega_0$ or
\begin{equation}
    \omega_0 = \sqrt{(\lambda v_0^2/\hbar)^2 + \Omega^2} - \lambda v_0^2/\hbar,
    \label{eq:demo_omega0}
\end{equation}
to ensure that the same part of the BCF is probed at the level of the Redfield equation.

Additionally, we require that the equilibrium expectation values of $(C_\Omega + C_\Omega^\dagger)^2$ match to align the coupling strength for the transition at $\Omega$.
Since $C_{\Omega_0}^{\rm HO} + [C_{\Omega_0}^{\rm HO}]^\dagger = v_0 q$, this condition becomes
\begin{equation}
    {\rm tr}_S \left[ \left( C_\Omega + C_\Omega^\dagger \right)^2  \rho_{S,{\rm eq}} \right] \simeq v_0^2 \braket{q^2}_{\rm eq},
    \label{eq:demo_v0}
\end{equation}
where $\rho_{S,{\rm eq}}$ is a state with relevant equilibrium properties.

Using \erefs{eq:demo_omega0} and (\ref{eq:demo_v0}), we determine $(\omega_0,v_0)$ and evaluate the error $\delta_\Omega$ in the harmonic oscillator system associated with the transition at $\Omega$.
In general systems, multiple transitions contribute to the dynamics simultaneously.
To maintain the simplicitiy of the testing procedure, we treat each transition independently and evaluate $\delta_\Omega$ for each $\Omega$.
We then estimate the total error in the worst-case scenario, assuming all errors add without cancellation:
\begin{equation}
    \delta_{\rm HO} \equiv \sum_{\Omega \geq 0} p(\Omega) \delta_\Omega,
    \label{eq:demo_total_error}
\end{equation}
where $p(\Omega)$ is the weight of how much the transition at $\Omega$ contributes to the dynamics.
We take $p(\Omega)$ to be proportional to the coupling strength,
\begin{equation*}
    p(\Omega) \propto {\rm tr}_S \left[ \left( C_\Omega + C_\Omega^\dagger \right)^2  \rho_{S,{\rm eq}} \right].
\end{equation*}

Three remarks on the proposed procedure are in order:
\begin{enumerate}[label=(\arabic*)]
    \item Ideally, $\rho_{S,{\rm eq}} = \lim_{t \to \infty} \rho_S(t)$, but evaluating the right-hand side requires a reliable $L_{\rm mod}(t)$, which this procedure aims to test.
    For practical evaluation, we instead take
    \begin{equation}
        \rho_{S,{\rm eq}} \propto \exp(-\beta H_{S, {\rm eff}}),
        \label{eq:demo_rhoSeq_Gibbs}
    \end{equation}
    with $H_{S, {\rm eff}} = H_S - \lambda V_S^2$ [see \eref{eq:method_CT}], which is the equilibrium state in the classical limit.
    \item For high-dimensional systems, the large number of distinct $\Omega$ makes evaluating \eref{eq:demo_total_error} impractical. We therefore include only the contributions accounting for up to 99 $\%$ of the total in the followings.
    \item According to \eref{eq:demo_omega0}, $\Omega = 0$ implies $\omega_0 = 0$, a case beyond the scope of our framework. We therefore restrict ourselves to cases where the $\Omega = 0$ contribution is negligible.
\end{enumerate}

In closing, we note that HEOM can exhibit instabilities, particularly in the strong coupling regime, even with a sufficiently large hierarchy depth \cite{Dunn19,Krug23}.
One possible source is the pathological behavior of $L_{\rm mod}(t)$. For instance, the Redfield decay rate is given by ${\rm Re}[ \Gamma(\Omega) ] = 2 \mathcal{F}[L](\Omega)$, and instability arises in the weak coupling regime if $\mathcal{F}[L_{\rm mod}]
 \ll 0$.
Such cases can be detected within our procedure by testing the stability of the surrogate harmonic oscillator dynamics.
Instability may also occur even with an accurate $L_{\rm mod}(t)$ due to the inherent sensitivity of the non-normal HEOM generator to small numerical errors \cite{Dunn19,MT24}.
This issue may be alleviated by modifying the truncation scheme or adopting alternative representations, such as the pseudomode form [\eref{eq:analysis_IPansatz}] or a continuous coordinate basis \cite{MT24,TN22}.
This sensitivity depends on the details of the system, and it cannot be detected within our approach.

\subsection{Examples}
\label{demo_example}

The construction of the testing procedure is guided primarily by intuitive arguments. In this section, we evaluate its effectiveness by applying it to a two-spin system and a transmon-resonator system.
In what follows, we focus on the relative error $\delta$.
For quantitative purposes, we take $\delta < 0.01$ to indicate a sufficiently small error for practical use.
Computational details are given in \aref{app:num}.

\subsubsection{Two-spin system}
\label{demo_example_twospin}

As an example, we consider a system of two spins with
\begin{equation*}
\begin{gathered}
    H_S / \hbar = \frac{\omega^{(1)}_s}{2} \sigma_z^{(1)} + \frac{\omega^{(2)}_s}{2} \sigma_z^{(2)} + g \, \sigma_x^{(1)} \sigma_x^{(2)}, \\
    V_S = \sigma_x^{(1)} + \sigma_x^{(2)},
\end{gathered}
\end{equation*}
where $\sigma_i^{(n)}$ $(i = x,y,z)$ are the Pauli matrices for the $n$th spin ($n=1,2$), $\omega^{(n)}_s$ the corresponding energy splittings, and $g$ the spin-spin coupling strength.
We take $J_{\rm exp}(\omega)$ given by \eref{eq:analysis_Jexp} and set $\omega_s^{(1)} = 1.2$, $\omega_s^{(2)} = 0.8$, $g = 0.4$, $s = 1$, $\alpha = 0.2$, and $\omega_c = 10$ in units of $\hbar = \beta = 1$.
The asymmetry of the spins $\omega^{(1)}_s \ne \omega^{(2)}_s$ ensures a unique steady state.

\begin{table}[h]
  \centering
  \begin{tabular}{|c|c|c|c|} \hline
    $\Omega$ & $\omega_0$ & $v_0$ & $p(\Omega)$ \\ \hline
    0.630 & 0.306 & 0.704 & 0.647 \\ \hline
    1.524 & 0.933 & 0.883 & 0.353 \\ \hline
  \end{tabular}
  \caption{
  Harmonic oscillator parameters and transition weights for the two-spin system.
  }
  \label{tab:demo_2spin}
\end{table}

The coupling $V_S$ induces two transitions with Bohr frequencies $\Omega = 0.630$ and $1.524$, corresponding to flips of spin 2 and 1, respectively.
Following \sref{demo_test}, we determine the harmonic oscillator parameters $(\omega_0, v_0)$ and the transition weights $p(\Omega)$ (Table \ref{tab:demo_2spin}), and use \eref{eq:demo_total_error} to estimate the total error for testing $L_{\rm mod}(t)$.

\begin{figure}[t]
  \includegraphics[keepaspectratio, scale=0.29]{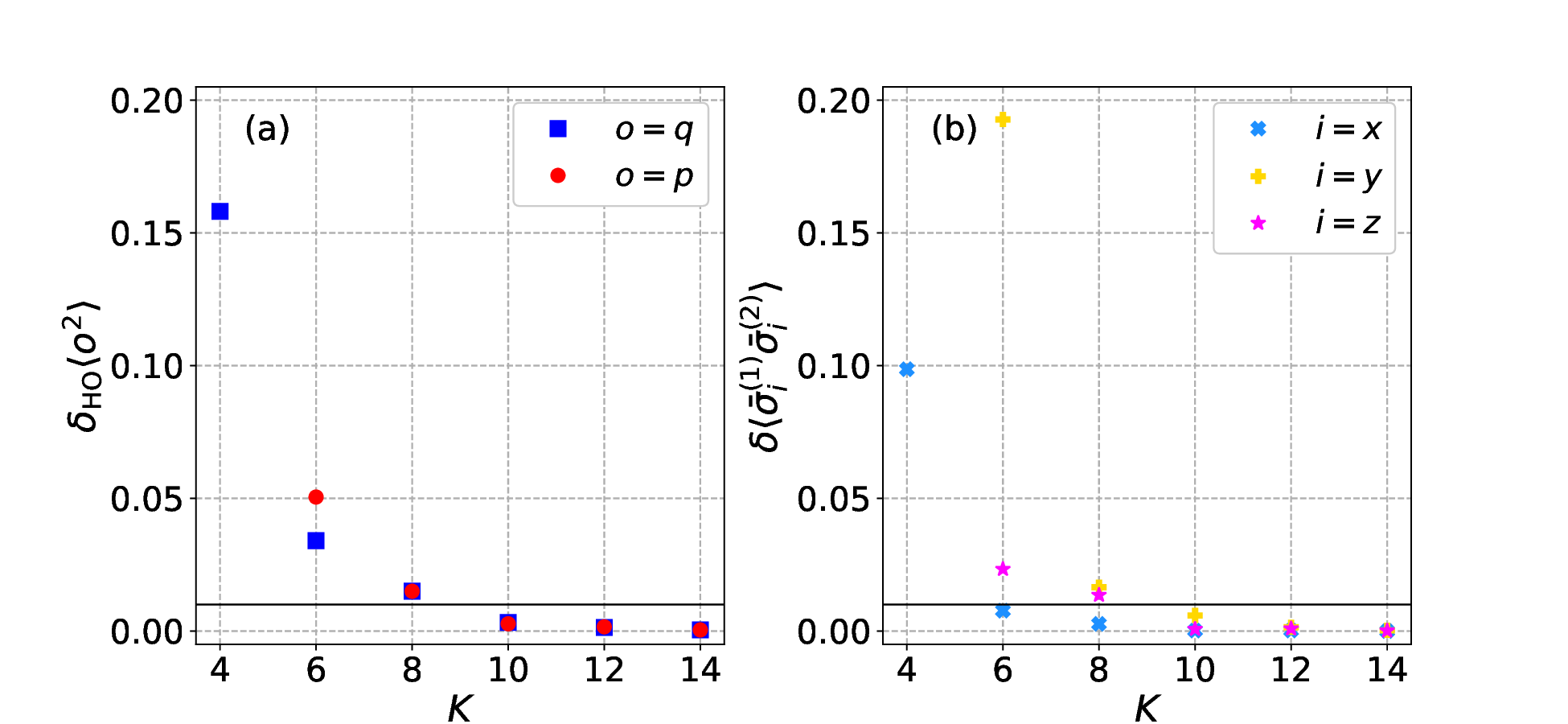}
  \caption{
  Error in the equilibrium expectation value vs $K$ for (a) the harmonic oscillator [\eref{eq:demo_dHOo2}] and (b) the two-spin system [\eref{eq:demo_dsisi}].
  In both panels, the horizontal black line marks $y = 0.01$, where $y$ denotes the plotted vertical-axis quantity.
  }
  \label{fig:demo_2spin_exp}
\end{figure}

We examine the equilibrium expectation values.
Figure~\ref{fig:demo_2spin_exp}(a) shows the $K$ dependence of
\begin{equation}
    \delta_{\rm HO} \! \braket{o^2} \equiv \sum_{\Omega \geq 0} p(\Omega) \, \delta_\Omega \! \braket{o^2},
    \label{eq:demo_dHOo2}
\end{equation}
for $o=q,p$, where $\delta_\Omega \! \braket{o^2}$ is the relative error [\eref{eq:analysis_do2}] computed with the harmonic oscillator parameters for the transition at $\Omega$.
For $\Omega = 0.630$ and $1.524$, the $K$ dependences of $\delta_\Omega \! \braket{o^2}$ are similar (not shown).
The horizontal line in the figure marks $\delta_{\rm HO} \! \braket{o^2} = 0.01$, which is reached for $K \geq 10$.

For the two-spin system, we focus on equilibrium spin-spin correlations.
Defining $\braket{O}_K \equiv {\rm tr}[O \, \rho (t_f)]$ as the expectation value computed with $K$ exponential terms, the correlations are given by $\braket{\bar{\sigma}_i^{(1)} \! \bar{\sigma}_i^{(2)}}_{K} \ (i = x,y,z)$, where $\bar{\sigma}_i^{(n)} = \sigma_i^{(n)} - \braket{\sigma_i^{(n)}}_K \ (n = 1,2)$.
The relative error is
\begin{equation}
    \delta \! \braket{\bar{\sigma}_i^{(1)} \! \bar{\sigma}_i^{(2)}} \equiv \left| \frac{\braket{\bar{\sigma}_i^{(1)} \bar{\sigma}_i^{(2)}}_{K_{\rm ref}} - \braket{\bar{\sigma}_i^{(1)} \bar{\sigma}_i^{(2)}}_{K}}{\braket{\bar{\sigma}_i^{(1)} \bar{\sigma}_i^{(2)}}_{K_{\rm ref}}} \right|,
    \label{eq:demo_dsisi}
\end{equation}
with a reference value $K_{\rm ref} = 16$.

Figure~\ref{fig:demo_2spin_exp}(b) plots $\delta \! \braket{\bar{\sigma}_i^{(1)} \bar{\sigma}_i^{(2)}} \ (i=x,y,z)$.
The $K$ dependences in the two-spin case qualitatively agree with those of the harmonic oscillator, and the condition $\delta \! \braket{\bar{\sigma}_i^{(1)} \! \bar{\sigma}_i^{(2)}} < 0.01$ $(i = x, y, z)$ is satisfied for $K \geq 10$, consistent with \fref{fig:demo_2spin_exp}(a).
Similar consistency is found for equilibrium correlation functions (Appendix \ref{app:2spin}).

\subsubsection{Transmon-resonator system}
\label{demo_example_cqed}

As a more complex example, here we consider a system of circuit quantum electrodynamics consisting of a transmon and a resonator:
\begin{equation}
\begin{gathered}
    H_S / \hbar = \frac{\omega_t}{4} \left[ Y_t^2 - \frac{2}{\epsilon} \cos(\sqrt{\epsilon} X_t) \right]
    + \omega_r N_r
    + g Y_t Y_r, \\
    V_S = Y_r,
\end{gathered}
\label{eq:demo_cqedH}
\end{equation}
where $X_i = a_i + a_i^\dagger$, $Y_i = i (a_i^\dagger - a_i)$, and $N_i = a_i^\dagger a_i$ with annihilation (creation) operators $a_i$ ($a_i^\dagger$) for the transmon $(i = t)$ and resonator ($i = r$),
$\omega_t$ ($\omega_r$) the transmon (resonator) frequency, $\epsilon$ the transmon anharmonicity, and $g$ the transmon-resonator coupling strength.

We investigate two types of spectral density.
We start with $J_{\rm exp}(\omega)$ in \eref{eq:analysis_Jexp}.
In units $\hbar = \omega_t = 1$, we set $\omega_r = 2$, $g = 0.4$, $\epsilon = 0.15$, $s = 1$, $\alpha = 0.1$, $\omega_c = 5$, and $\beta = \infty$ (zero temperature).
Larger values of $g$ and $\alpha$ than in typical experiments are chosen to create a numerically challenging case.

\begin{table}[h]
  \centering
  \begin{tabular}{|c|c|c|c|} \hline
    $\Omega$ & $\omega_0$ & $v_0$ & $p(\Omega)$ \\ \hline
    2.08500 & 1.68817 & 1.33186 & 0.70446 \\ \hline
    0.76310 & 0.65289 & 0.69141 & 0.19645 \\ \hline
    2.09154 & 2.07711 & 0.24069 & 0.02418 \\ \hline
    2.09547 & 2.08149 & 0.23691 & 0.02343 \\ \hline
    2.08544 & 2.07598 & 0.19475 & 0.01584 \\ \hline
    2.07421 & 2.06673 & 0.17314 & 0.01252 \\ \hline
    2.09190 & 2.08759 & 0.13136 & 0.00721 \\ \hline
    0.76096 & 0.75719 & 0.12301 & 0.00633 \\ \hline
    0.78129 & 0.77787 & 0.11712 & 0.00576 \\ \hline
    0.74326 & 0.74096 & 0.09604 & 0.00385 \\ \hline
  \end{tabular}
  \caption{
  Harmonic oscillator parameters and transition weights for the transmon-resonator system with $J_{\rm exp}(\omega)$.
  }
  \label{tab:demo_cqed}
\end{table}

When $g = 0$, only transitions that change the resonator photon number by one contribute to the dynamics.
For $g = 0.4$, additional transitions become relevant, and we include the ten listed in Table \ref{tab:demo_cqed} for error estimation.

\begin{figure}[t]
  \includegraphics[keepaspectratio, scale=0.29]{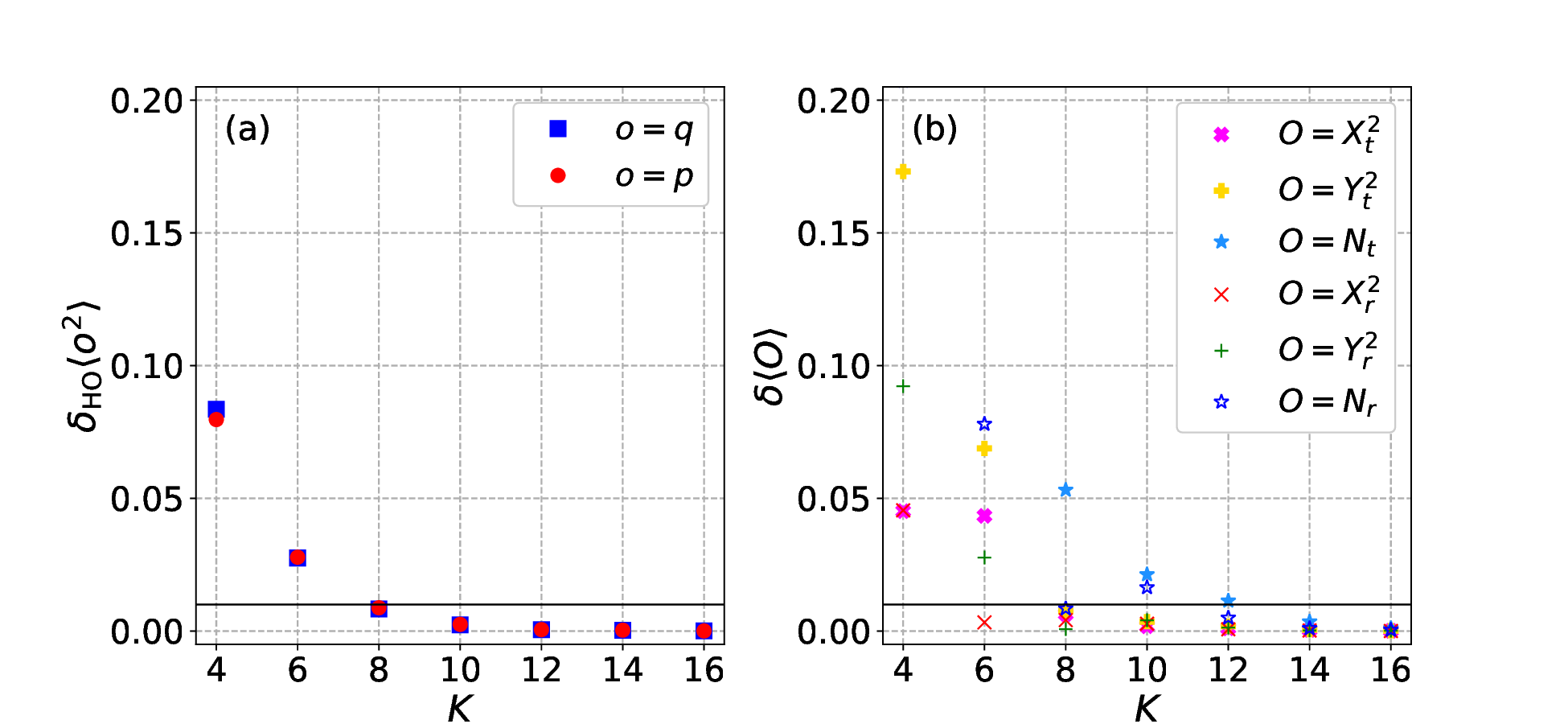}
  \caption{
  Error in the equilibrium expectation value vs $K$ for (a) the harmonic oscillator and (b) the transmon-resonator system with $J_{\rm exp}(\omega)$.
  In both panels, the horizontal black line marks $y = 0.01$, as in \fref{fig:demo_2spin_exp}.
  }
  \label{fig:demo_cqed_exp}
\end{figure}

We examine the equilibrium expectation values.
Figure~\ref{fig:demo_cqed_exp}(a) shows the $K$ dependence of $\delta_{\rm HO} \! \braket{o^2}$ ($o = q, p$) defined in \eref{eq:demo_dHOo2}.
The $K$ dependence of $\delta_\Omega \! \braket{o^2}$ are similarly for all $\Omega$ (not shown).
The condition $\delta_{\rm HO} \! \braket{o^2} < 0.01$, indicated by the horizontal black line, is reached for $K \geq 8$.

For the transmon-resonator system, we focus on $X_i^2$, $Y_i^2$, and $N_i$.
The relative error, evaluated as in the two-spin case [\eref{eq:demo_dsisi}] with $K_{\rm ref} = 18$, is shown in \fref{fig:demo_cqed_exp}(b).
Two behaviors are observed.
For $O = X_i^2$ and $Y_i^2$, $\delta \! \braket{O}$ decreases rapidly and falls below $0.01$ for $K \geq 8$, consistent with the harmonic oscillator case.
For $O = N_i$, on the other hand, the error decays more slowly, with $\delta \! \braket{N_i} < 0.01$ achieved only for $K \geq 14$.
This difference arises because $N_i$ probes different components of the density matrix, despite the identity $N_i = (X_i^2 + Y_i^2 - 2)/4$.

\begin{figure}[t]
  \includegraphics[keepaspectratio, scale=0.45]{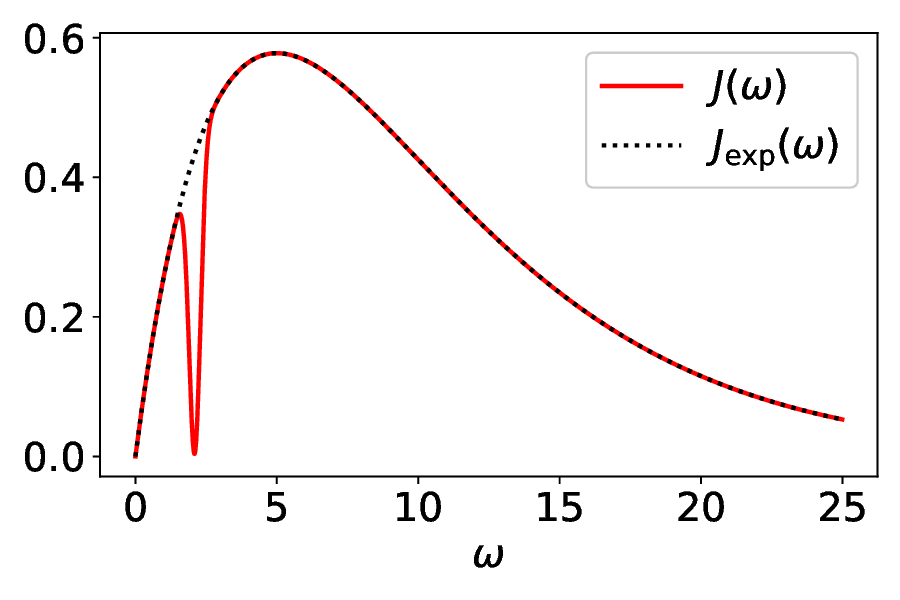}
  \caption{
  Spectral density with filter (red) and without filter (dotted black) [\eref{eq:demo_Jfilter}]. All quantities are shown in units of $\hbar = \omega_t = 1$.
  }
  \label{fig:demo_Jfilter}
\end{figure}

To test the procedure with a structured spectral density, we apply a filter to $J_{\rm exp}(\omega)$ with the same parameters as above.
In the Redfield equation, the decay rate is given by ${\rm Re}[ \Gamma(\Omega) ] = 2 \mathcal{F}[L](\Omega)$, implying that it can be suppressed by reducing $J(\Omega)$ \cite{Reed10}.
Following this idea, we take
\begin{equation}
    J(\omega \geq 0) = |J_{\rm exp}(\omega) - f(\omega)|,
    \label{eq:demo_Jfilter}
\end{equation}
with a filter function $f(\omega)$.
Here, we consider suppressing the decay of the resonator and set $f(\omega) = 0.99 \, J_{\rm exp}(\omega_{f}) \, \exp(-(\omega-\omega_f)^2 / (2 \sigma_f^2))$ with $\omega_f = 2.1$ and $\sigma_f = 0.2$, yielding $J(\omega_f) = 0.01 J_{\rm exp}(\omega_f)$ (see \fref{fig:demo_Jfilter}).
The harmonic oscillator parameters in this case are similar to Table \ref{tab:demo_cqed}, with slight deviations in $(\omega_0,v_0,p(\Omega))$ due to differences in $\lambda$ [see \eref{eq:demo_rhoSeq_Gibbs}].

\begin{figure}[t]
  \includegraphics[keepaspectratio, scale=0.38]{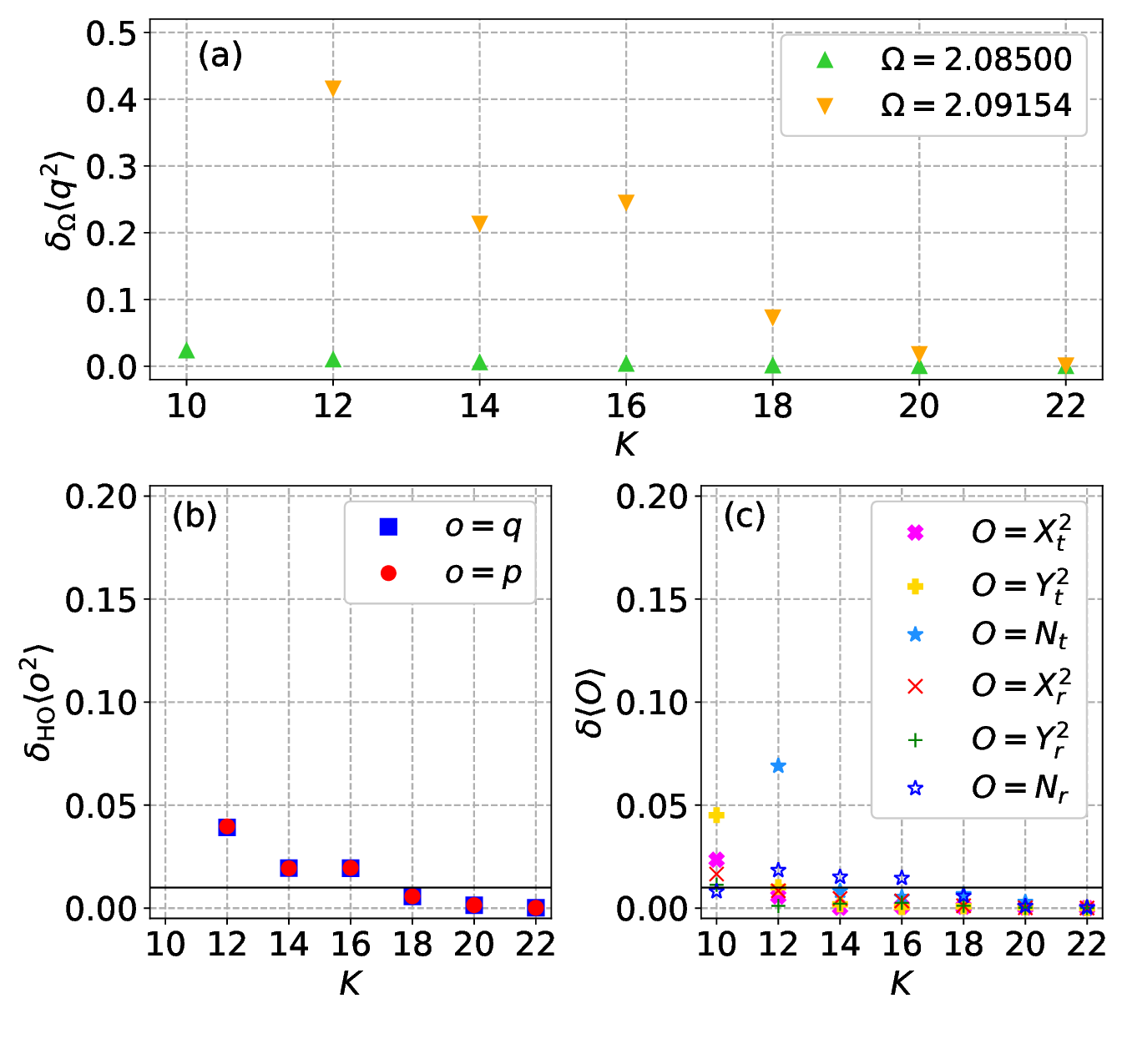}
  \caption{
  Error in the equilibrium expectation values $K$ for the transmon-resonator system with the filtered spectral density \eref{eq:demo_Jfilter}.
  Panel (a) shows $\delta_\Omega \! \braket{q^2}$ for two transitions $\Omega = 2.08500$ and $\Omega = 2.09154$.
  Panels (b) and (c) correspond to the same quantities as in \fref{fig:demo_cqed_exp}.
  }
  \label{fig:demo_cqed_filter_exp}
\end{figure}

In contrast with the previous examples, $\delta_\Omega \! \braket{o^2}$ exhibits two distinct behaviors depending on $\Omega$, as shown in \fref{fig:demo_cqed_filter_exp}(a).
For $\Omega \simeq 0.76$, the $K$ dependence matches that of $\Omega = 2.08500$ (top row in Table \ref{tab:demo_cqed}), with $\delta_\Omega \! \braket{o^2} < 0.01$ for $K \geq 14$.
For $\Omega \simeq 2.08$ (except the top row), on the other hand, the behavior follows that of $\Omega = 2.09154$ (third row in Table \ref{tab:demo_cqed}), and $\delta_\Omega \! \braket{o^2} < 0.01$ is achieved only for $K \geq 22$.
The latter cases directly probe the structured filter region because $\omega_0 \simeq \omega_f$ and $v_0$ is small, which might account for the larger error.

The total error for the harmonic oscillator systems is shown in \fref{fig:demo_cqed_filter_exp}(b).
We see that $\delta_{\rm HO} \! \braket{o^2} < 0.01$ is reached for $K \geq 18$.
The relative error for the transmon-resonator system ($K_{\rm ref} = 24$) is shown in \fref{fig:demo_cqed_filter_exp}(c).
Here, $\delta \! \braket{O} < 0.01$ is achieved for $K \geq 14$ when $O = X_i^2, Y_i^2$, and for $K \geq 18$ when $O = N_i$.
The latter is consistent with the harmonic oscillator estimate.

In summary, these examples support the validity of our proposed testing procedure.
Although accurately capturing the quantity dependence of the error is challenging, the surrogate harmonic oscillator systems reproduce the overall error trends of the original dynamics.
This correspondence allows us to use the harmonic oscillator results as a guide for choosing a reasonable $L_{\rm mod}(t)$.

\section{Concluding remarks}
\label{conclusion}

To test the accuracy of dynamics computed using a model BCF, we proposed using a harmonic oscillator system with known exact solutions.
We presented several methods to numerically evaluate these solutions for general spectral densities (\sref{method_test}) and proposed a moment-based representation to simplify the computation (\sref{method_moment}).
Applying the testing procedure to a two-spin system and a transmon-resonator system, we found that the harmonic oscillator results provide reasonable estimates of the simulation accuracy (\sref{demo}).

Using the methodologies introduced in this article, we evaluated the performance of several state-of-the-art methods for efficiently constructing BCFs.
For the Ohmic spectral density, we found that ESPRIT proposed in \rcite{Takahashi24} achieves the smallest error (\sref{analysis_error}).
However, we also identified its potential limitation when applied to sub-Ohmic spectral densities at finite temperatures, where AAA proposed in \rcite{Xu22} more accurately captures the characteristic low-frequency divergence of the equilibrium correlation function (\sref{analysis_sub}).

In addition to the main findings, this study opens up several promising directions for future research.
One such direction is the application of the moment representation introduced in \sref{method_moment} to systems involving anharmonic oscillators.
Although the truncation is no longer exact in such cases, this representation may still offer computational advantages over the conventional Fock basis, as suggested by a study of an optomechanical system with weak anharmonicity \cite{MT24} and by the demonstrated efficiency of HEOM in treating general systems.
Another potential direction involves the model BCF given in \eref{eq:method_Lmod}, which is compatible with the more general form $v_k(t) = t^{n_k} e^{-z_k t}$, where $n_k \in \mathbb{Z}_{\geq 0}$ and $z_k \in \mathbb{C}$.
The exponential ansatz \eref{eq:analysis_Lmod_exp} corresponds to the special case $n_k = 0$.
This generalized form has been shown to be effective for describing low-temperature baths \cite{Cui19,Zhang20}.
The analysis in \sref{analysis_error} implies that methods reducing the problem to linear optimization tend to outperform brute-force nonlinear optimization approaches.
These insights motivate the development of a new method capable of efficiently handling the cases with $n_k > 0$.

\begin{acknowledgments}
    I thank Yoshitaka Tanimura for his valuable input throughout the research.
    I am also grateful to Shoki Koyanagi, Jianshu Cao, Michael Thoss, and Johannes Feist for fruitful discussions.
    This work was supported by JSPS KAKENHI Grant No. JP23KJ1157.
\end{acknowledgments}

\section*{Data Availability}
The data that support the findings of this article are openly available \cite{Data}.

\appendix
\section{Pseudomode method}
\label{app:IP}

In this appendix, we discuss the pseudomode method, where the open quantum dynamics [\eref{eq:method_rhoS}] are described by embedding the system into an extended space with auxiliary bosonic modes (pseudomodes).
While early works \cite{Imamoglu94, Garraway97, Tamascelli17, Smirne22} assumed that the extended state evolves under a Gorini-Kossakowski-Sudarshan-Lindblad (GKSL) equation \cite{GKS, Lindblad}, the framework remains applicable even when the evolution is not completely positive \cite{Xu23,Park24} or Hermitian-preserving \cite{Lambert19, Luo23, Menczel24}.
As noted in \rcite{Garraway97}, non-GKSL master equations offer greater flexibility in modeling BCFs.

As recently shown in \rcites{Link22,Xu23,MT24}, the key difference between the pseudomode method and the HEOM is the use of a moment-based representation (see \aref{app:moment} for details) for the auxiliary state. To illustrate this, consider a GKSL master equation with $Q$ pseudomodes:
\begin{equation}
\begin{gathered}
    \frac{d}{dt} \varrho(t) = \\
    - \frac{i}{\hbar} \left( H_S + \sum_{k,k'=1}^Q \hbar \omega_{k,k'} a_k^\dagger \! a_{k'} + V_S X_{ps} \right)^\times \varrho(t) \\
    + \sum_{k=1}^Q \kappa_k \mathcal{D}[a_k] \, (\varrho(t)),
\end{gathered}
\label{eq:IP_ME}
\end{equation}
where $\varrho(t)$ is the density operator in the extended space, $a_k$ ($a_k^\dagger$) is the annihilation (creation) operator for the $k$th pseudomode, the interaction operator $X_{ps}$ is assumed to be $X_{ps} = \sum_{k = 1}^Q g_k (a_k + a_k^\dagger)$, and $\mathcal{D}[a_k] (\varrho) = a_k \varrho a_k^\dagger - (a_k^\dagger a_k \varrho + \varrho a_k^\dagger a_k)/ 2$ is the dissipator.
The fitting parameters are $\omega_{k,k'} (= \omega_{k',k}) \in \mathbb{R}$, $g_k \in \mathbb{R}$, and $\kappa_k \in \mathbb{R}_{> 0}$.
The internal dynamics of the pseudomodes is governed by $\mathfrak{L}_{ps} = - i \sum_{k,k'=1}^Q \omega_{k,k'} (a_k^\dagger a_{k'})^\times + \sum_{k=1}^Q \kappa_k \mathcal{D}[a_k]$, where the vacuum state $\ket{\bm{0}}$ is a steady state, $\mathfrak{L}_{ps} (\ketbra{\bm{0}}{\bm{0}}) = 0$.
Assuming the initial state $\varrho(0) = \rho_S(0) \ketbra{\bm{0}}{\bm{0}}$, the model BCF corresponding to \eref{eq:IP_ME} reads \cite{Tamascelli17}
\begin{gather}
    L_{\rm mod}(t \geq 0) \equiv \frac{1}{\hbar} {\rm tr} \left[ X_{ps} e^{\mathfrak{L}_{ps} t} (X_{ps} \ketbra{\bm{0}}{\bm{0}}) \right] \nonumber \\
    = \frac{1}{\hbar} \bm{g}^\top e^{- i \tilde{\omega} t} \bm{g}, \label{eq:IP_Lmod}
\end{gather}
with $\bm{g}^\top = [g_1, \dots, g_Q]$
and $\tilde{\omega}_{k,k'} = \omega_{k,k'} - (i/2) \delta_{k,k'} \kappa_k$.

In \aref{app:moment}, we explore the moment representation of the oscillator system state.
On the other hand, here we consider the moment representation of the pseudomode state.
In this case, the map representing the transformation, \eref{eq:moment_T}, is defined as \cite{Xu23}
\begin{equation*}
    \mathcal{S} \equiv e^{\sum_{k=1}^Q a_k^L (a_k^\dagger)^R},
\end{equation*}
which operates trivially on the system state space.
Let $\varphi(t) \equiv \mathcal{S} (\varrho(t))$.
With the aid of \erefs{eq:moment_op}, operating $\mathcal{S}$ on \eref{eq:IP_ME} yields the evolution equation for $\varphi(t)$ as
\begin{equation}
\begin{gathered}
    \frac{d}{dt} \varphi(t) = - \frac{i}{\hbar} H_S^\times \varphi(t) \\
    -i \sum_{k,k'=1}^Q \tilde{\omega}_{k,k'} a_k^\dagger a_{k'} \varphi(t) + i \sum_{k,k'=1}^Q \tilde{\omega}^*_{k,k'} \varphi(t) a_k^\dagger a_{k'} \\
    - \frac{i}{\hbar} \sum_{k=1}^Q g_k \left[ V_S^L (a_k^\dagger \varphi(t)) - V_S^R (\varphi(t) a_k) \right] \\
    - \frac{i}{\hbar} \sum_{k=1}^Q g_k V_S^\times \left( a_k \varphi(t) + \varphi(t) a_k^\dagger \right).
\end{gathered}
\label{eq:IP_HEOMop}
\end{equation}
Reducing the pseudomode degrees of freedom by taking the matrix element as $\varphi_{\bm{m},\bm{n}}(t) \equiv \braket{\bm{m}|\varphi(t)|\bm{n}}$, we find
\begin{equation*}
\begin{gathered}
    \frac{d}{dt} \varphi_{\bm{m},\bm{n}}(t) = - \frac{i}{\hbar} H_S^\times \varphi_{\bm{m},\bm{n}}(t) \\
    -i \sum_{k,k'=1}^Q \tilde{\omega}_{k,k'} \sqrt{m_k (m_{k'}+1-\delta_{k,k'})} \ \varphi_{\bm{m}-\bm{e}_k + \bm{e}_{k'},\bm{n}}(t) \\
    + i \sum_{k,k'=1}^Q \tilde{\omega}^*_{k,k'} \sqrt{n_k(n_{k'}+1-\delta_{k,k'})} \ \varphi_{\bm{m},\bm{n}-\bm{e}_k + \bm{e}_{k'}}(t) \\
    - \frac{i}{\hbar} \sum_{k=1}^Q g_k \left[ \sqrt{m_k} \, V_S^L \varphi_{\bm{m}-\bm{e}_k, \bm{n}}(t) - \sqrt{n_k} \, V_S^R \varphi_{\bm{m},\bm{n}-\bm{e}_k}(t) \right] \\
    - \frac{i}{\hbar} \sum_{k=1}^Q g_k V_S^\times \left( \sqrt{m_k + 1} \, \varphi_{\bm{m}+\bm{e}_k,\bm{n}}(t) \right. \\
    \left. + \sqrt{n_k + 1} \, \varphi_{\bm{m},\bm{n}+\bm{e}_k}(t) \right).
\end{gathered}
\end{equation*}
These equations are in fact a special case of the HEOM [\eref{eq:method_HEOM}] with
\begin{equation*}
    \bm{\theta} =
    i \left[
    \begin{array}{c}
    \bm{g} \\ \hline
    \bm{g}
    \end{array}
    \right], \ \
    \bm{v}(0) =
   - \frac{i}{\hbar} \left[
    \begin{array}{c}
    \bm{g} \\ \hline
    \bm{g}
    \end{array}
    \right],
\end{equation*}
\begin{equation*}
    D = \left[
    \begin{array}{c|c}
    I & O \ \\ \hline
    O & O
    \end{array}
    \right], \ \
    \bar{D} = \left[
    \begin{array}{c|c}
    O & O \ \\ \hline
    O & I
    \end{array}
    \right],
\end{equation*}
and
\begin{equation*}
    Z = \left[
    \begin{array}{c|c}
     i \tilde{\omega} & O \\ \hline
    O & - i \tilde{\omega}^*
    \end{array}
    \right],
\end{equation*}
with the $Q \times Q$ identity matrix $I$ and zero matrix $O$.
From \eref{eq:method_Lmod}, the model BCF corresponding to these is given by
\begin{equation*}
    L_{\rm mod}(t) = \frac{1}{\hbar}
    \left[
    \begin{array}{c}
    \bm{g} \\ \hline
    \bm{0}
    \end{array}
    \right]^\top
    \left[
    \begin{array}{c|c}
     e^{- i \tilde{\omega} t} & O \\ \hline
    O & e^{i \tilde{\omega}^* t}
    \end{array}
    \right]
    \left[
    \begin{array}{c}
    \bm{g} \\ \hline
    \bm{g}
    \end{array}
    \right],
\end{equation*}
which is consistent with \eref{eq:IP_Lmod}.

In summary, we show that the HEOM can be derived by expressing the pseudomode equation in the moment representation.
Note that \eref{eq:IP_HEOMop}, which presents the operator form of the HEOM, offers a bosonic particle picture of the auxiliary quantities. This picture was introduced in the literature under the name {\it dissipaton} \cite{Yan14,Yan16}.

In \sref{analysis_fit}, we introduce the interacting pseudomode approach to a model BCF.
The ansatz \eref{eq:analysis_IPansatz} is obtained by setting $\bm{g} = \sqrt{\hbar/(2 \pi)} \ \bm{l}$ in \eref{eq:IP_Lmod}.
In \rcite{Mascherpa20}, the interaction between the pseudomodes is restricted to the nearest-neighbor ones for efficient simulations using the tensor-network state (see also \rcite{Somoza19}).
The fully interacting pseudomodes were considered in \rcites{Medina21,Lednev24} and we focus on this approach as it offers greater flexibility in modeling BCFs.
In \sref{analysis_fit}, we use the open source package available in \rcite{SDF} to determine the parameters. The relative tolerance is set $5 \times 10^{-5}$.
Regarding the initial guess, we use the AAA results, $L_{\rm mod}(t) = \sum_{k=1}^K d_k \exp(-z_k t)$, and set the parameters as follows: $\omega_{k,k'} = \delta_{k,k'} {\rm Im}(z_k)$, $\kappa_k = 2 {\rm Re}(z_k)$, and $l_k = \sqrt{2 \pi D_k}$, where $D_k = {\rm Re}(d_k)$ if ${\rm Re}(d_k) \geq 0$ and $D_k = 10^{-5}$ otherwise.

\section{Exact solutions for the harmonic oscillator system}
\label{app:exact}

In this appendix, we present the explicit formulas of the exact solutions for the harmonic oscillator system.

\subsection{Solution of the Heisenberg equation of motion}
\label{exact_EOM}

It is instructive to start with the solution of the Heisenberg equation of motion.
With \erefs{eq:method_Htot}, (\ref{eq:method_CT}), and (\ref{eq:method_HO}), we find that $q(t) \equiv \exp(i H^\times t/\hbar) (q)$ satisfies
\begin{equation*}
\begin{gathered}
    \frac{d^2}{dt^2} q(t) + \frac{1}{M} \int_0^t d\tau \, \eta(t-\tau) \frac{d}{d\tau}q(\tau) + \omega_0^2 q(t) \\
    = - \frac{1}{M v_0} X^0_B(t) - \frac{\eta(t)}{M} q,
\end{gathered}
\end{equation*}
with $X^0_B(t) = \exp(i H_B^\times t/\hbar) (X_B)$,
$M = \hbar/ (\omega_0 v_0^2)$,
and $\eta(t) = (2 / \pi) \int_0^\infty d\omega (J(\omega) / \omega) \cos(\omega t)$.
We see from this equation of motion that $\omega_0$ represents the effective frequency of the oscillator and $\eta(t)$ represents the friction coefficient.
This equation can be solved using, for instance, the Laplace transform \cite{FRH11}.
The general solution is given by
\begin{equation}
\begin{gathered}
    q(t) = q \frac{d}{dt} G_+(t) + p \, \omega_0 G_+(t) \\
    - \frac{1}{M v_0} \int_0^t d\tau \, G_+ (t-\tau) X^0_B(\tau),
\end{gathered}
\label{eq:exact_q}
\end{equation}
where $G_+(t)$, which is defined in the Laplace space as
\begin{gather}
    \hat{G}_+(s) \equiv \int_0^\infty dt \, G_+(t) e^{- s t} \nonumber \\
    = \frac{1}{\omega_0^2 + s^2 + s \hat{\eta}(s)/M}, \label{eq:exact_Ghat_+}
\end{gather}
with $\hat{\eta}(s) = \int_0^\infty dt \, \eta(t) e^{- s t}$,
is the solution of the differential equation
\begin{equation*}
    \frac{d^2}{dt^2} G_+(t) + \frac{1}{M} \int_0^t d\tau \, \eta(t-\tau) \frac{d}{d\tau} G_+(\tau) + \omega_0^2 G_+ (t) = 0,
\end{equation*}
with initial conditions $G_+(0) = 0$ and $(d/dt) G_+(0) = 1$.

We can use \eref{eq:exact_q} to obtain the exact expectation values.
For instance, with the initial state $\rho_{\rm tot}(0) = \rho_S(0) \rho_B$, the expectation value of $q^2$ is given by
\begin{equation*}
\begin{gathered}
    {\rm tr}_S [q^2 \rho_S(t)] = \braket{q^2}_0 \left[ \frac{dG_+}{dt} (t) \right]^2
    + \braket{p^2}_0 \omega_0^2 \left[ G_+ (t) \right]^2 \\
    + \braket{qp+pq}_0 \omega_0  G_+ (t) \frac{d G_+}{dt} (t) \\
    + \frac{\omega_0}{M} \int_0^t d\tau \int_0^t d\tau' \, G_+(\tau) L(\tau-\tau') G_+(\tau'),
\end{gathered}
\end{equation*}
where $\braket{o}_0 \equiv {\rm tr}_S [o \, \rho_S(0)]$ is the initial expectation value of an operator $o$.

\subsection{Analytic solutions}
\label{app:exact_solutions}

Using the path integral formulation, the authors of \rcite{GSI88} found analytic expressions of the equilibrium expectation values $\braket{o^2}_{\rm eq}$ and autocorrelation functions $C_{oo}(t)$ for $o=q,p$ (see \sref{method_test} for their definitions).
First, the equilibrium expectation values are given by
\begin{equation*}
    \braket{q^2}_{\rm eq} = \frac{1}{\beta \hbar \omega_0} \sum_{n = -\infty}^\infty \frac{\omega_0^2}{\omega_0^2 + \nu_n^2 + \zeta_n},
\end{equation*}
and
\begin{equation*}
    \braket{p^2}_{\rm eq} = \frac{1}{\beta \hbar \omega_0} \sum_{n = -\infty}^\infty \frac{\omega_0^2 + \zeta_n}{\omega_0^2 + \nu_n^2 + \zeta_n},
\end{equation*}
with  $\nu_n = 2 \pi n / (\beta \hbar)$ and $\zeta_n = \nu_n \hat{\eta}(\nu_n) / M$.
Second, the equilibrium autocorrelation functions in the frequency domain are given by
\begin{equation*}
\begin{gathered}
    \mathcal{F}[C_{qq}](\omega) = \frac{2 \omega_0}{1 - e^{- \beta \hbar \omega}} {\rm Im} [\hat{G}_+ (- i \omega)]
\end{gathered}
\end{equation*}
with $\hat{G}_+ (-i \omega) = [\omega_0^2 - \omega^2 - i\omega \hat{\eta} (- i \omega) / M]^{-1}$ [see \eref{eq:exact_Ghat_+}],
and
\begin{equation*}
    \mathcal{F}[C_{pp}](\omega) = \left( \frac{\omega}{\omega_0} \right)^2 \mathcal{F}[C_{qq}](\omega).
\end{equation*}
Note that the zero temperature limit can be obtained by replacing the infinite sum with the integral for $\braket{o^2}_{\rm eq}$ and the factor $1/(1-\exp(-\beta \hbar \omega))$ with the step function for $\mathcal{F}[C_{oo}](\omega)$.

We examine the behavior of $\mathcal{F}[C_{qq}](\omega)$ in the vicinity of $\omega = 0$ for different Ohmicities.
Inserting \eref{eq:method_eta-iomega} into \eref{eq:exact_Ghat_+} yields
\begin{equation*}
\begin{gathered}
    {\rm Im} [\hat{G}_+ (- i \omega)] = \\
    \frac{J(\omega)/M}{(\omega^2 - \omega_0^2 + \omega \, {\rm Im}[\hat{\eta}(-i\omega)]/M)^2 + (J(\omega)/M)^2  },
\end{gathered}
\end{equation*}
with ${\rm Im}[\hat{\eta}(-i\omega)]$ involves the Cauchy principal value.
Assuming $\lim_{\omega \to 0^+} \omega \, {\rm Im}[\hat{\eta}(-i\omega)] = 0$, which is numerically confirmed for the cases discussed in this article, the denominator becomes $\omega_0^4$ in the limit $\omega \to 0^+$. This leads to $\lim_{\omega \to 0^+} {\rm Im}[\hat{G}_+(-i\omega)] = J(\omega) / (M \omega_0^4)$.
Assuming $J(\omega \simeq 0^+) = \mathcal{J} \, \omega^s$ near $\omega = 0$, we thus find at finite temperatures
\begin{equation*}
    \mathcal{F}[C_{qq}](\omega \simeq 0^+) = \frac{2 \mathcal{J}}{\beta} \left( \frac{v_0}{\hbar \omega_0} \right)^2 \omega^{s-1}.
\end{equation*}
This indicates that $\mathcal{F}[C_{qq}](\omega)$ in the vicinity of $\omega = 0$ is sensitive to the Ohmicity as
\begin{equation}
    \lim_{\omega \to 0} \mathcal{F}[C_{qq}](\omega)
    \begin{cases}
    = \infty & (0 < s < 1), \\
    > 0 & (s = 1), \\
    = 0 & (s > 1).
    \end{cases}
    \label{eq:exact_FCqq_near_0}
\end{equation}
Similar behavior is expected for different systems.
Recently, \rcite{BZFP24} proposed a machine learning approach to infer Ohmicity from the time evolution of a system observable.
However, this approach may be impractical as it relies on assumptions, such as an initially factorized state, that are difficult to justify experimentally.
Given that linear response functions like $\mathcal{F}[C_{qq}](\omega)$ are experimentally accessible, \eref{eq:exact_FCqq_near_0} might offer a more practical alternative for identifying Ohmicity.

\subsection{Cases with analytic $\hat{\eta}(s)$}
\label{app:exact_analytic_eta}

The above formulas indicate that we only need $\hat{\eta}(|s|) \ (|s| > 0)$ to evaluate the equilibrium expectation values and $\hat{\eta}(- i \omega) \ (\omega \in \mathbb{R})$ for the equilibrium autocorrelation functions.
Combining these two, we need $\hat{\eta}(s) \ ({\rm Re}(s) \geq 0)$ to evaluate the exact solutions.
Here, we present analytic expressions of this quantity for several spectral densities.

One example is the generalized Meier-Tannor form [\eref{eq:GMT_J}] (see \aref{app:GMT} for details).
In this case, $\eta(t)$ is given by a sum of exponential functions [see \eref{eq:method_GMT}] and the integral for the Laplace transform $\hat{\eta}(s) = \int_0^\infty dt \, \eta(t) \exp(-s t)$ yields
\begin{equation}
    \hat{\eta}(s) = - 2i \sum_j \left[ \frac{c_j}{\mu_j (s + \mu_j)} - \frac{c_j^*}{\mu_j^* (s + \mu_j^*)} \right],
    \label{eq:exact_eta_GMT}
\end{equation}
for ${\rm Re}(s) \geq 0$.

The other example is
the exponential cutoff \eref{eq:analysis_Jexp}, where the BCF is given for arbitrary parameters by
\begin{equation*}
\begin{gathered}
  L(t) = \frac{\alpha \omega_c^2}{2 (\beta \hbar \omega_c)^{s+1}} \Gamma(s+1) \\
  \times \left[ \zeta(s+1, z^*(t)) + \zeta(s+1, z(t)+1) \right],
\end{gathered}
\end{equation*}
with the Gamma function $\Gamma(s)$, the Hurwitz zeta function $\zeta(s,a)$, and $z(t) = (1 - i \omega_c t) / (\beta \hbar \omega_c)$.
In the Ohmic case $s = 1$, the analytic expression of $\hat{\eta}(s)$ is available.
Using the formulas in \erefs{eq:method_etas} and (\ref{eq:method_eta-iomega}), we find
\begin{equation}
    \hat{\eta}(|s|) = - \alpha \, {\rm Im} \left[ E(i |s| / \omega_c) e^{i |s| / \omega_c}  \right],
    \label{eq:exact_eta_exp_positive}
\end{equation}
for $|s| > 0$, and
\begin{equation}
\begin{gathered}
    \hat{\eta}(- i \omega) = \frac{\pi}{2} \alpha \, e^{- |\omega| / \omega_c} \\
    + i \frac{\alpha}{2} \left[ E(\omega/\omega_c) e^{\omega/\omega_c} - E(-\omega/\omega_c) e^{-\omega/\omega_c} \right].
\end{gathered}
    \label{eq:exact_eta_exp_imaginary}
\end{equation}
for $\omega \in \mathbb{R}$, where $E(z)$ is the exponential integral defined by
\begin{equation*}
    E(z) = {\rm p.v.} \! \int_0^\infty dx \frac{e^{-(x + z)}}{x + z}.
\end{equation*}
This can be evaluated, for instance, using the scipy functions scipy.linalg.exp1 for $z \in \mathbb{C} \land z \notin \mathbb{R}_{\leq 0}$ and
scipy.linalg.expi for $z \in \mathbb{R}_{< 0}$.

\section{Generalized Meier-Tannor decomposition}
\label{app:GMT}

In this appendix, we explore the generalized Meier-Tannor decomposition introduced in \eref{eq:method_GMT}.
The spectral density is given by the Fourier transform of the imaginary part, $J(\omega) \equiv i \mathcal{F}[{\rm Im}(L)](\omega)$, and reads
\begin{equation}
    J(\omega) = \sum_j \frac{  4 \omega \, {\rm Im}(c_j (\omega^2 + (\mu_j^*)^2))}{\left[ (\omega+\Omega_j)^2+\Gamma_j^2 \right] \left[ (\omega-\Omega_j)^2+\Gamma_j^2 \right]}.
    \label{eq:GMT_J}
\end{equation}
Using this, we can derive ${\rm Re}[L(t)]$ as
\begin{equation}
\begin{gathered}
    {\rm Re}[L(t \geq 0)] = 2 \sum_j {\rm Im} \left[ c_j \cot \left( \frac{\beta \hbar \mu_j}{2} \right) e^{- \mu_j t} \right] \\
    + \frac{2}{\beta \hbar} \sum_{n = 1}^\infty i J(i \nu_n) e^{- \nu_n t},
\end{gathered}
\label{eq:GMT_ReL}
\end{equation}
with $\nu_n = 2 \pi n / (\beta \hbar)$.
The terms in the second line represent the Matsubara contribution.

In \sref{method_test}, we propose leveraging this decomposition to accurately evaluate the exact values of $\braket{o^2}_{\rm eq}$ and $\mathcal{F}[C_{oo}](\omega)$.
These quantities can be computed from the complex parameters $\{ c_j, \mu_j \}_j$ using \eref{eq:exact_eta_GMT}.
One way to obtain the parameters $\{ c_j, \mu_j \}_j$ is by fitting ${\rm Im}[L(t \geq 0)]$ with the ESPRIT algorithm (see \sref{analysis_fit}).
Although the algorithm does not inherently enforce the constraints ${\rm Im}[L(t)] \in \mathbb{R}$ and $J(\omega \geq 0) \geq 0$, we expect these conditions to be satisfied if the fitting of ${\rm Im}[L(t)]$ is sufficiently accurate.

In \sref{analysis_fit}, we introduce the method of GMT$\&$FIT to construct a model BCF $L_{\rm mod}(t)$.
This two-step approach involves fitting $J(\omega)$ and ${\rm Re}[L(t)]$.
In \sref{analysis_error}, we apply ESPRIT to fit ${\rm Im}[L(t \geq 0)]$, which yields $J(\omega)$ as \eref{eq:GMT_J}.
The resulting complex parameters $\{ c_j, \mu_j \}_j$ define ${\rm Re}[L(t)]$ as \eref{eq:GMT_ReL}, which includes the infinite Matsubara sum.
To approximate the Matsubara contribution using a finite number of real exponentials, we adopt the ansatz
\begin{equation*}
\begin{gathered}
    {\rm Re}[L_{\rm mod} (t \geq 0)] = 2 \sum_j {\rm Im} \left[ c_j \cot \left( \frac{\beta \hbar \mu_j}{2} \right) e^{- \mu_j t} \right] \\
    + \sum_n b^{\rm MT}_n e^{- \gamma^{\rm MT}_n t},
\end{gathered}
\end{equation*}
with real fitting parameters $\{ b^{\rm MT}_n, \gamma^{\rm MT}_n \}_n$ constrained by $\gamma^{\rm MT}_n > 0$.
We determine these using the scipy function scipy.optimize.minimize with the "L-BFGS-M" method.
The initial guess is set to $b^{\rm MT}_n = 2 i J(i \nu_n)/(\beta \hbar)$ and $\gamma^{\rm MT}_n = \nu_n$ motivated by \eref{eq:GMT_ReL}.

\section{Moment representation}
\label{app:moment}

\subsection{Introduction}

In this appendix, we explore the moment representation of the oscillator system state.
The moment representation of $\rho_{\bm{j}}$ (omitting the time argument for now) is introduced in \eref{eq:method_moment}:
\begin{equation}
  \phi_{m,n,\bm{j}} = \frac{{\rm tr}_S(a^m \rho_{\bm{j}} (a^\dagger)^n )}{\sqrt{m! \, n!}}.
\label{eq:moment_def}
\end{equation}
Let $\phi_{\bm{j}} \equiv \sum_{m,n = 0}^{\infty} \phi_{m,n,\bm{j}} \ketbra{m}{n}$, with which $\{ \phi_{m,n,\bm{j}}\}_{m,n=0}^\infty$ can be obtained from its matrix elements as $\phi_{m,n,\bm{j}} = \braket{m|\phi_{\bm{j}}|n}$.
The map $\mathcal{S}$ transforming $\rho_{\bm{j}}$ to $\phi_{\bm{j}}$, $\phi_{\bm{j}} \equiv \mathcal{S}(\rho_{\bm{j}})$, is linear and is given by
\begin{equation}
    \mathcal{S} = \sum_{n=0}^\infty \frac{\left[ a^L (a^\dagger)^R \right]^n}{n!} = e^{a^L (a^\dagger)^R},
    \label{eq:moment_T}
\end{equation}
where the superscripts $L$ and $R$ represent the operation from the left and right, respectively, as introduced below \eref{eq:method_M}.
This indicates that the inverse $\mathcal{S}^{-1}$, which describes the inverse transformation $\rho_{\bm{j}} =\mathcal{S}^{-1} (\phi_{\bm{j}})$, is given by $\mathcal{S}^{-1} = \exp(-a^L (a^\dagger)^R)$.
We can show
\begin{equation}
  {\rm tr}_S (o) = \braket{0|\mathcal{S} (o) |0},
  \label{eq:moment_trace}
\end{equation}
and
\begin{equation}
  \begin{gathered}
    \mathcal{S} (a o) = a \mathcal{S}(o), \ \
    \mathcal{S} (a^\dagger \! o) = a^\dagger \mathcal{S}(o) + \mathcal{S}(o) a^\dagger, \\
    \mathcal{S} (o a^\dagger) = \mathcal{S}(o) a^\dagger, \ \
    \mathcal{S} (o a) = a \mathcal{S}(o) + \mathcal{S}(o) a,
  \end{gathered}
  \label{eq:moment_op}
\end{equation}
for any system operator $o$.

\subsection{HEOM in the moment representation}
\label{app:moment_HEOM}

The expression of the HEOM [\eref{eq:method_HEOM}] in the moment representation can be derived by applying $\mathcal{S}$ and using \erefs{eq:moment_op}.
Inserting \erefs{eq:method_CT} and (\ref{eq:method_HO}), we find
\begin{equation}
\begin{gathered}
    \frac{d}{dt} \phi_{\bm{j}} (t)
    = - \frac{i}{\hbar} \mathcal{S}[H_S^\times \rho_{\bm{j}} (t)] - \sum_{k,k'=1}^{K} Z_{k,k'}^{\bm{j}} \phi_{\bm{j}-\bm{e}_k+\bm{e}_{k'}} (t) \\
    + v_0 \sum_{k=1}^{K} \sqrt{j_k} \left( [D \bm{v}(0)]_k \mathcal{S}[q \rho_{\bm{j}-\bm{e}_k}] - [\bar{D} \bm{v}(0)]_k \mathcal{S}[\rho_{\bm{j}-\bm{e}_k} q] \right) \\
    - \frac{v_0}{\hbar} \sum_{k=1}^{K} \theta_k \sqrt{j_k+1} \, \mathcal{S}[q^\times \rho_{\bm{j}+\bm{e}_k}],
\end{gathered}
\label{eq:moment_HEOM}
\end{equation}
with
\begin{equation*}
\begin{gathered}
    \mathcal{S}[H_S^\times \rho_{\bm{j}} (t)] = (\hbar \omega_0 + \lambda v_0^2) (a^\dagger \! a)^\times \phi_{\bm{j}}(t) \\
    + \lambda v_0^2 \left[ \frac{1}{2} (a^\dagger)^2 \phi_{\bm{j}}(t) - \frac{1}{2} \phi_{\bm{j}}(t) a^2 + a^\dagger \phi_{\bm{j}}(t) a^\dagger - a \, \phi_{\bm{j}}(t) a \right],
\end{gathered}
\end{equation*}
\begin{equation*}
    \mathcal{S}[q \rho_{\bm{j}-\bm{e}_k} (t)] =
    q \phi_{\bm{j}-\bm{e}_k} (t) + \frac{1}{\sqrt{2}} \phi_{\bm{j}-\bm{e}_k} (t) a^\dagger,
\end{equation*}
\begin{equation*}
    \mathcal{S}[\rho_{\bm{j}-\bm{e}_k} (t) q] =
    \phi_{\bm{j}-\bm{e}_k} (t) q + \frac{1}{\sqrt{2}} a \, \phi_{\bm{j}-\bm{e}_k} (t),
\end{equation*}
and
\begin{equation*}
    \mathcal{S}[q^\times \rho_{\bm{j}+\bm{e}_k} (t)] =
    \frac{1}{\sqrt{2}} \left( a^\dagger \phi_{\bm{j}+\bm{e}_k} (t) - \phi_{\bm{j}+\bm{e}_k} (t) a \right).
\end{equation*}
Taking the matrix element, we see that $(d/dt) \phi_{m,n,\bm{j}} (t)$ with depth $\mathcal{H} \equiv m + n + \sum_{k=1}^K j_k$ depends only on elements $\phi_{m',n',\bm{j}'}  (t)$ with depth $m' + n' + \sum_{k=1}^K j_k' = \mathcal{H}$ or $\mathcal{H}-2$.
For example, the last line of \eref{eq:moment_HEOM} involves $\phi_{m-1,n,\bm{j}+\bm{e}_k} (t)$ and $\phi_{m,n-1,\bm{j}+\bm{e}_k} (t)$ for $k = 1,2,\dots,K$, both with depth $\mathcal{H}$.
This structure ensures that elements $\{ \phi_{m,n,\bm{j}} \}$ with depth $\leq \mathcal{H}$ are decoupled from those with greater depth.
This enables exact truncation in computing their evolution.

\subsection{Computation of $\braket{o^2}_{\rm mod}$ and $C^{\rm mod}_{oo}(t)$}
\label{app:moment_computation}

Here, we present the way of computing $\braket{o^2}_{\rm mod}$ [\eref{eq:analysis_omod}] and $C^{\rm mod}_{oo} (t)$ [\eref{eq:analysis_Coomod}] within the HEOM framework in the original and moment representations.
To this end, we introduce the orthonormal vector notation of the auxiliary states as $\rho \equiv \sum_{\bm{j}} \rho_{\bm{j}} |\bm{j})$, with $(\bm{j}|\bm{j}') = \delta_{\bm{j},\bm{j}'}$.
The HEOM [\eref{eq:method_HEOM}] can then be expressed as $(d/dt) \rho(t) = \mathfrak{L}(\rho(t))$ with the linear generator $\mathfrak{L}$.
Due to the linearity, its formal solution is given by $\rho(t) = \exp(\mathfrak{L} t) (\rho(0))$, where the initial state reads $\rho(0) = \rho_S(0) |\bm{0})$ [see the discussion below \eref{eq:method_HEOM}].
We can then evaluate the quantities as
\begin{equation*}
    \braket{o^2}_{\rm mod} = {\rm tr}_S \left[ (\bm{0}| (o^2)^L e^{\mathfrak{L} t_f} (\rho(0))   \right]
\end{equation*}
and
\begin{equation}
    C^{\rm mod}_{oo}(t) = {\rm tr}_S \left[ (\bm{0}| o^L e^{\mathfrak{L} t} \left( o^L e^{\mathfrak{L} t_f} (\rho(0)) \right) \right].
    \label{eq:moment_Coo}
\end{equation}

A similar expression in the moment representation can be obtained using $\phi = \mathcal{S}(\rho)$, where $\mathcal{S}$ operates trivially on the auxiliary state space.
The evolution equation for $\rho(t)$ yields $(d/dt)\phi(t) = \mathfrak{M}(\phi(t))$ with  $\mathfrak{M} = \mathcal{S} \mathfrak{L} \mathcal{S}^{-1}$, which corresponds to the HEOM in the moment representation [\eref{eq:moment_HEOM}].
The initial state is given by $\phi(0) = \mathcal{S}(\rho_S(0)) |\bm{0})$. With the initial vacuum state $\rho_S(0) = \ketbra{0}{0}$, for instance, we have $\mathcal{S}(\rho_S(0)) = \ketbra{0}{0}$.

Using \eref{eq:moment_trace}, the above formulas for $\braket{o^2}_{\rm mod}$ and $C^{\rm mod}_{oo}(t)$ can be transformed to the moment representation expression as
\begin{equation}
    \braket{o^2}_{\rm mod} = \braket{0| (\bm{0}| \underline{(o^2)^L} \, e^{\mathfrak{M} t_f} (\phi(0)) |0}
    \label{eq:moment_o2_mr}
\end{equation}
and
\begin{equation}
    C^{\rm mod}_{oo}(t) =
    \braket{0| (\bm{0}| \underline{o^L} \, e^{\mathfrak{M} t} \left( \underline{o^L} \, e^{\mathfrak{M} t_f} (\phi(0)) \right) |0}.
    \label{eq:moment_Coo_mr}
\end{equation}
In these equations, we have introduced $\underline{\mathcal{O}} \equiv \mathcal{S} \mathcal{O} \mathcal{S}^{-1}$ for $\mathcal{O} = (o^2)^L$ and $o^L$, where the explicit expressions can be found using \erefs{eq:moment_op}.
These are the second-order moments of the ladder operators.
Therefore, the truncation at the depth $\mathcal{H} = 2$ (setting $\phi_{m,n,\bm{j}} = 0$ for $m + n + \sum_{k=1}^K j_k > 2$) is sufficient to compute them.

Throughout this article, we evaluate
\erefs{eq:moment_o2_mr} and (\ref{eq:moment_Coo_mr}) by computing the matrix exponentiation $\exp(\mathfrak{M} t)$ using the scipy function scipy.sparse.linalg.expm to avoid time discretization errors.

\section{Computational details}
\label{app:num}

This appendix summarizes the computational procedures used in the main text.

\subsection{Section \ref{analysis_error}}
\label{app:num_analysis_error}

To construct model BCFs, data points are equidistantly sampled for both time- and frequency-domain fitting methods.
For the time-domain methods (ESPRIT and GMT$\&$FIT), we use $\Delta t_{\rm data} = 0.01$ over $t \in [0, t_{\rm max})$ with $t_{\rm max} = 20$.
For the frequency-domain methods (AAA and IP), we use $\Delta \omega_{\rm data} = 0.1$ over $\omega \in [-\omega_{\rm max}, \omega_{\rm max})$ with $\omega_{\rm max} = 300$.
In GMT$\&$FIT, the spectral density is fitted with eight exponential terms, and the remaining $K-8$ terms account for the Matsubara contribution.

Throughout this article, the HEOMs [\eref{eq:analysis_HEOM}] for the harmonic oscillator system are solved by computing the matrix exponential of the generator in the moment representation (see \aref{app:moment_computation} for details).
Starting from the vacuum state $\rho_S(0) = \ketbra{0}{0}$, the system reaches a steady state at $t_f = 30$.

The Fourier transform of the correlation function [\eref{eq:analysis_Coomod}] is computed as
$\mathcal{F}[C^{\rm mod}_{oo}] (\omega) = 2 {\rm Re}[ \int_{0}^{t_f} dt \, C^{\rm mod}_{oo}(t) \exp(i \omega t)]$, where the integral is evaluated using Simpson's rule with a time step of $0.1$.

\subsection{Section \ref{analysis_sub}}
\label{app:num_analysis_sub}

\subsubsection{Construction of model BCFs}

For ESPRIT, data points are equidistantly sampled with $\Delta t_{\rm data} = 0.01$ over $t \in [0, t_{\rm max})$, and we set $t_{\rm max} = 200$.

\begin{figure}[t]
  \includegraphics[keepaspectratio, scale=0.4]{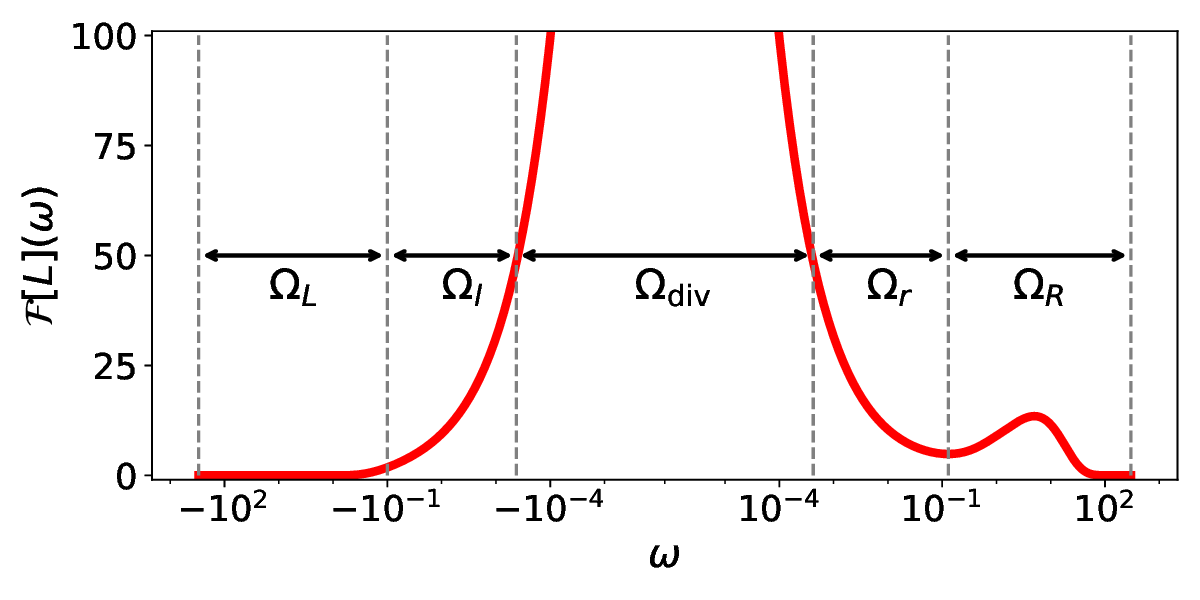}
  \caption{
  Partitioning of the frequency space into five subregions for frequency-domain fitting.}
  \label{fig:sub_subohmic_frequency_fit}
\end{figure}

For AAA, the fitting must capture both the divergent behavior near $\omega = 0$ and the smooth variation at large $|\omega|$.
To this end, we partition the frequency domain $\Omega \equiv [-\omega_{\rm max}, \omega_{\rm max})$ (with $\omega_{\rm max} = 300$) into five subregions as shown in \fref{fig:sub_subohmic_frequency_fit}.
The outer regions $\Omega_L$ and $\Omega_R$ capture smooth tails, the intermediate regions $\Omega_l$ and $\Omega_r$ resolve the steep rise near $\omega = 0$, and the central region $\Omega_{\rm div}$ (containing the divergence) is excluded from fitting.

In $\Omega_L$ and $\Omega_R$, data points are equidistantly sampled with $\Delta \omega_{\rm data} = 0.1$.
In $\Omega_l$ and $\Omega_r$, we adopt a finer logarithmic grid.
Let $\Omega_r \equiv [\omega_r^{(0)}, \omega_r^{(1)}] \ (0 < \omega_r^{(0)} < \omega_r^{(1)})$ and $\Omega_l \equiv [\omega_l^{(1)}, \omega_l^{(0)}] \ (\omega_l^{(1)} < \omega_l^{(0)} < 0)$.
In each region $\Omega_s$ ($s = r$ and $l$), we sample $N$ points $\{ \omega_{s,n} \}_{n=0}^{M-1}$ as
\begin{equation*}
    \log|\omega_{s,n}| = \log|\omega^{(0)}_{s}| + \frac{n}{N-1} \left( \log|\omega^{(1)}_{s}| - \log|\omega^{(0)}_{s}| \right).
\end{equation*}
We set $N = 100$, which is one order of magnitude smaller than the number of data points in the smooth regions.
The endpoints $\omega^{(0)}_s$ are chosen such that $\mathcal{F}[L](\omega) \leq 50$ in the fitting region. This yields $\omega^{(0)}_r = 4.2161 \times 10^{-4}$ and $\omega^{(0)}_l = - 4.1881 \times 10^{-4}$.
The inner boundaries of the smooth regions $\omega^{(1)}_s$ are arbitrarily set to $\omega^{(1)}_r = 0.13$ and $\omega^{(1)}_l = - 0.1$.

\subsubsection{Thermalization simulations}

To obtain the exact solutions, $\hat{\eta}(s) \ ({\rm Re}(s) \geq 0)$ is numerically computed using \eref{eq:method_etas} for $s \in \mathbb{R}_{> 0}$ and \eref{eq:method_eta-iomega} for ${\rm Re}(s) = 0$, where the $\omega'$ integral is performed up to $\omega' = 10^3$ with the scipy function ${\rm scipy.integrate.quad}$.
The quantities $\braket{o^2}_{\rm mod}$ and $\mathcal{F}[C_{oo}^{\rm mod}](\omega)$ are evaluated as described above.
Convergent results are obtained for $o = q$ with $t_f = 10^4$ (ESPRIT) and $t_f = 2\times10^5$ (AAA).

\subsection{Section \ref{demo_example_twospin}}
\label{app:num_demo_example_twospin}

Model BCFs are constructed using ESPRIT from data sampled as above with $\Delta t_{\rm data} = 0.01$ and $t_{\rm max} = 20$.

The two-spin dynamics are computed by solving the HEOM [\eref{eq:analysis_HEOM}] using the fourth-order Runge-Kutta method with step size $dt = 0.005$ and truncation condition $\rho_{\bm{j}}(t) = 0$ for $\sum_{k=1}^{K} j_k > 6$.
Starting from $\rho_S(0) \propto \exp(- \beta H_{S,{\rm eff}})$, the system reaches steady state at $t_f = 300$.

\subsection{Section \ref{demo_example_cqed}}
\label{app:num_demo_example_cqed}

Model BCFs are obtained via ESPRIT, using data sampled as above with $\Delta t_{\rm data} = 0.02$ and $t_{\rm max} = 40$.

The HEOMs for the transmon-resonator system are solved with step size $dt = 0.005$ and truncation condition $\rho_{\bm{j}}(t) = 0$ for $\sum_{k=1}^{K} j_k > 4$.
We employ the product Fock basis $\ket{n_t m_r} \ (N_i \ket{n_i} = n \ket{n_i})$ ($n, m = 0,1,2,3, 4$), giving a Hilbert-space dimension of 25.
We confirm that the population of $\ket{4_i}$ $(i = t,r)$ remains on the order of $10^{-4}$.
Starting from $\rho_S(0) \propto \exp(- \beta H_{S,{\rm eff}})$, the system reaches steady state at $t_f = 150$.

To evaluate $\hat{\eta}(s)$ for the filtered case, we follow the procedure discussed below \eref{eq:method_GMT}, where we fit ${\rm Im}[L(t \geq 0)]$ using ESPRIT.

\section{Fitting-range dependence of ESPRIT results in the sub-Ohmic case}
\label{app:subESPRIT}

\begin{figure}[t]
  \includegraphics[keepaspectratio, scale=0.38]{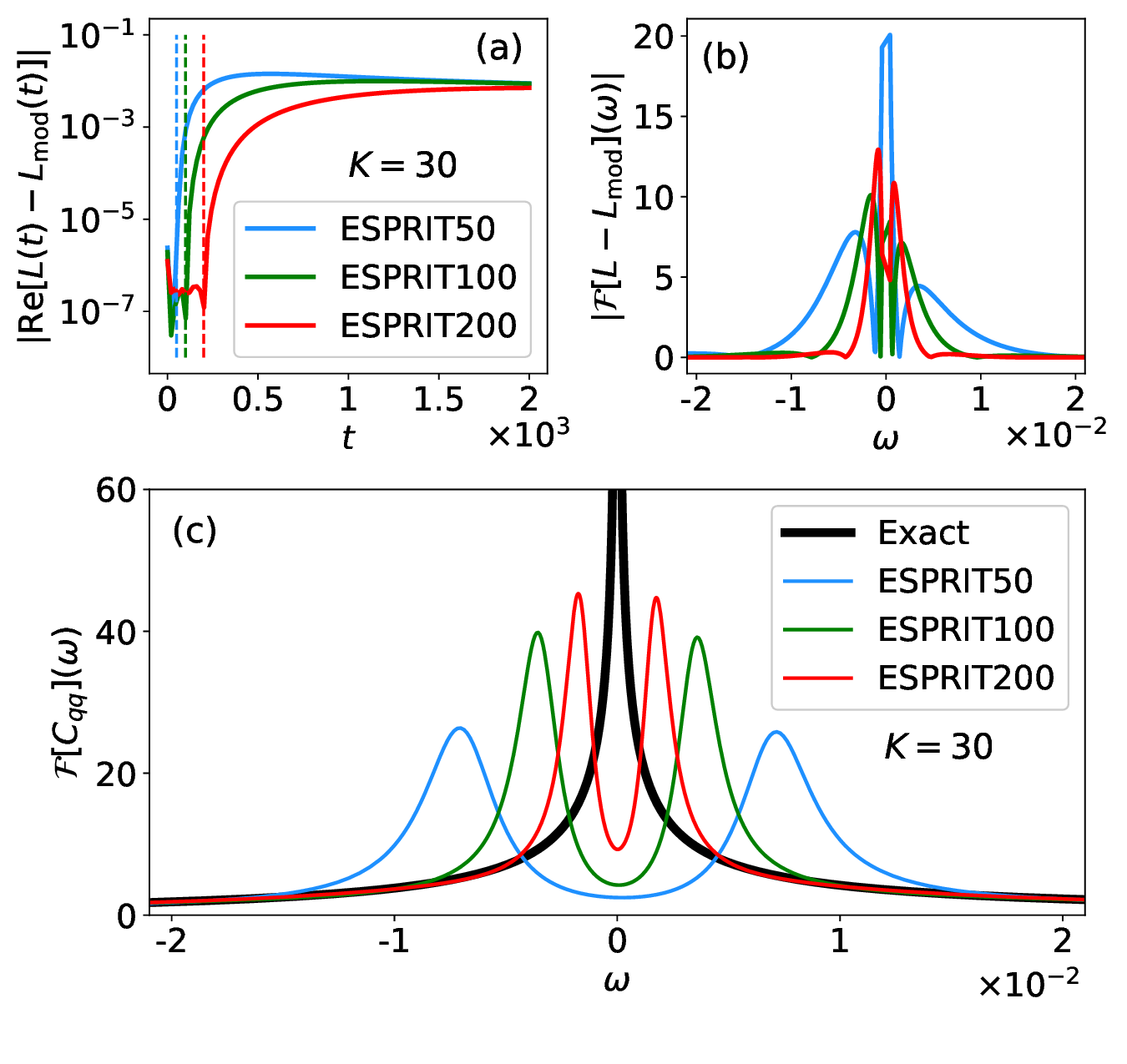}
  \caption{
  Dependence of ESPRIT fitting quality with $K = 30$ on $t_{\rm max}$.
  All quantities are shown in units of $\hbar = \omega_0 = v_0 = 1$.
  Error in (a) ${\rm Re}[L_{\rm mod}(t)]$ and (b) $\mathcal{F}[L_{\rm mod}](\omega)$ for $t_{\rm max} = 50$ (blue curve), $100$ (green curve), and $200$ (red curve).
  In panel (a), the thin dashed vertical lines show $t = t_{\rm max}$.
  In panel (b), the divergent region near $\omega = 0$, where $\mathcal{F}[L](\omega) > 50$, is excluded.
  (c) Comparison between the exact autocorrelation function (thick black curve) and the ESPRIT results (thin colored curves).
  }
  \label{fig:sub_subohmic_L_corr_fourier}
\end{figure}

In this appendix, we discuss how the ESPRIT results for the sub-Ohmic case depend on $t_{\rm max}$ by comparing cases with $t_{\rm max} = 50$, $100$, and $200$ (the analysis in \sref{analysis_sub} uses $t_{\rm max} = 200$).
All computations use $K = 30$, and we verify that larger $K$ values do not significantly affect the results.
Figure~\ref{fig:sub_subohmic_L_corr_fourier}(a) shows the error in ${\rm Re}[L_{\rm mod}](t)$ in the long-time region.
As expected, we see that the accuracy improves with increasing $t_{\rm max}$.
Figure~\ref{fig:sub_subohmic_L_corr_fourier}(b) shows the error in $\mathcal{F}[L_{\rm mod}](\omega)$ near $\omega = 0$.
Although the largest error does not decrease monotonically with increasing $t_{\rm max}$, the frequency range captured accurately becomes narrower as $t_{\rm max}$ increases.

We examine the reproducibility of the autocorrelation function.
First, we confirm that all values of $t_{\rm max}$ show good agreement on the scale of \fref{fig:analysis_subohmic_corr_fourier}(a) (not shown).
Figure~\ref{fig:sub_subohmic_L_corr_fourier}(c) compares the exact autocorrelation function with the ESPRIT results near $\omega = 0$.
Consistent with the improvements seen in \frefs{fig:sub_subohmic_L_corr_fourier}(a) and \ref{fig:sub_subohmic_L_corr_fourier}(b), larger $t_{\rm max}$ values more accurately capture the divergent behavior near $\omega = 0$.
Notably, for each value of $t_{\rm max}$, $\mathcal{F}[C_{qq}(\omega)]$ is accurately captured in the frequency range, where $|\mathcal{F}[L-L_{\rm mod}](\omega)|$ nearly vanishes [see \fref{fig:sub_subohmic_L_corr_fourier}(b)].

\section{Equilibrium correlation function errors in the two-spin system}
\label{app:2spin}

\begin{figure}[t]
  \includegraphics[keepaspectratio, scale=0.29]{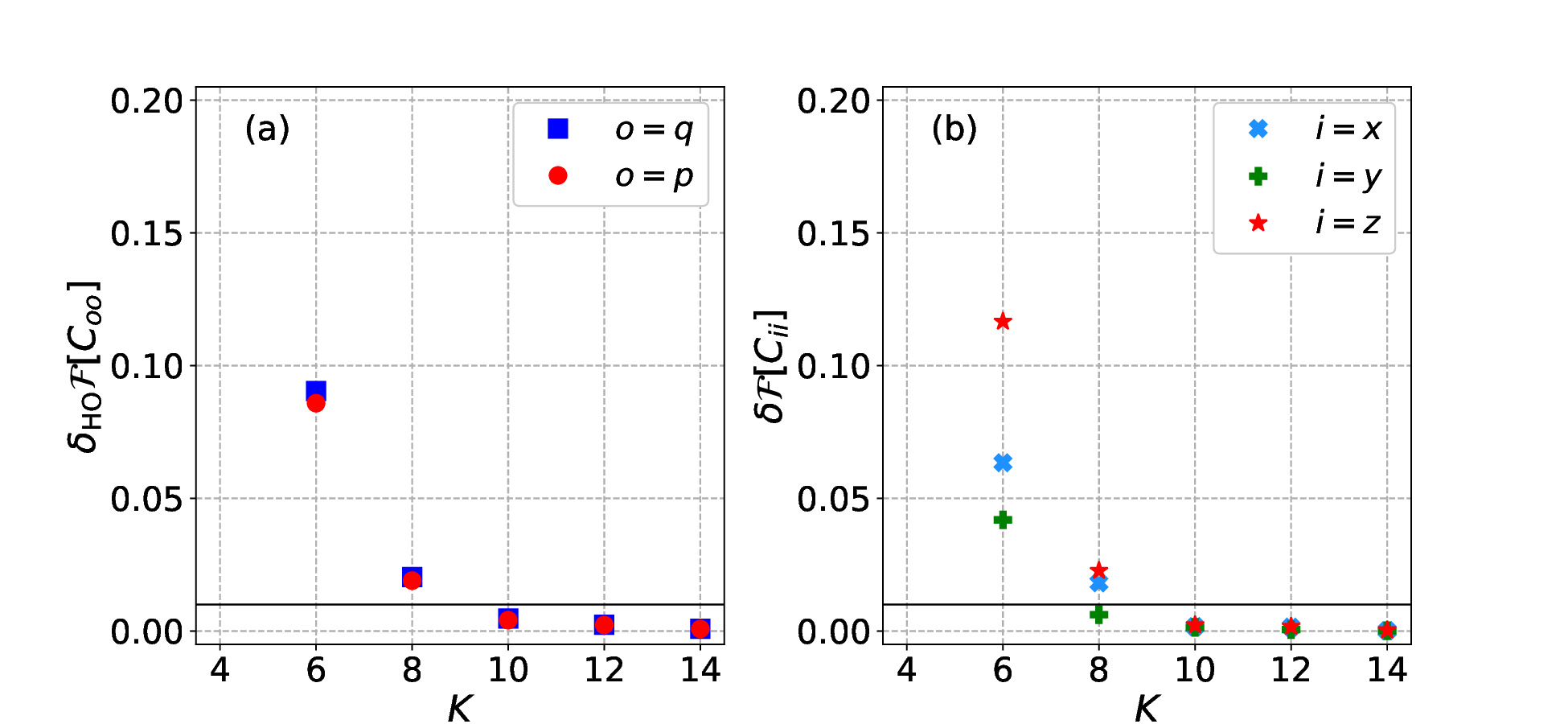}
  \caption{
  Error in the equilibrium correlation functions vs $K$ for (a) the harmonic oscillator [\eref{eq:demo_dHOFCoo}] and (b) the two-spin system [\eref{eq:demo_dFCii}].
  In both panels, the horizontal black line indicates $y = 0.01$, as in \fref{fig:demo_2spin_exp}.
  }
  \label{fig:demo_2spin_corr_fourier}
\end{figure}

In this appendix, we test the equilibrium correlation functions for the two-spin system in \sref{demo_example}.

For the harmonic oscillator, we define the relative error as
\begin{equation*}
    \delta \mathcal{F}[C_{oo}] \equiv \frac{\int_{-\infty}^\infty d\omega |\mathcal{F}[C_{oo}-C_{oo}^{\rm mod}](\omega)|}{\int_{-\infty}^\infty d\omega |\mathcal{F}[C_{oo}](\omega)|}.
\end{equation*}
Letting $\delta_\Omega \mathcal{F}[C_{oo}]$ denote this error computed with harmonic oscillator parameters for the transition at $\Omega$, the total error is
\begin{equation}
    \delta_{\rm HO} \mathcal{F}[C_{oo}] \equiv  \sum_{\Omega \geq0} p(\Omega) \, \delta_\Omega \mathcal{F}[C_{oo}].
    \label{eq:demo_dHOFCoo}
\end{equation}

For the two-spin system, we consider dynamical correlation functions $C^K_{ii}(t) \equiv \braket{\sigma_i^{(1)} \exp(iH_{\rm tot}^\times t/\hbar)(\sigma_i^{(2)})}_K$.
Within the HEOM framework, this is computed using \eref{eq:moment_Coo}, with $C^K_{ii}(t < 0) = {\rm tr} [\sigma_i^{(1)} \exp(-iH_{\rm tot}^\times t/\hbar) (\sigma_i^{(2)} \rho(t_f))]$, derived under the assumption that $\rho(t_f)$ is an equilibrium state.
As in the harmonic oscillator case, the relative error is quantified in the the frequency domain as
\begin{equation}
    \delta \mathcal{F}[C_{ii}] \equiv \frac{\int_{-\infty}^\infty d\omega \, |\mathcal{F}[C^{K_{\rm ref}}_{ii} - C^K_{ii}](\omega)|}{\int_{-\infty}^\infty d\omega \, |\mathcal{F}[C^{K_{\rm ref}}_{ii}](\omega)|},
    \label{eq:demo_dFCii}
\end{equation}
with $K_{\rm ref} = 16$.

Figures~\ref{fig:demo_2spin_corr_fourier}(a) and \ref{fig:demo_2spin_corr_fourier}(b) compare the $K$ dependence of these errors for the harmonic oscillator and two-spin cases.
As with the equilibrium expectation values, the two systems exhibit consistent behavior, with quantitative agreement that the errors fall below $0.01$ for $K \geq 10$ in both cases.

\end{document}